\documentclass[fleqn,usenatbib]{mnras}

\usepackage{newtxtext,newtxmath}
\usepackage[T1]{fontenc}
\usepackage{dsfont}
\usepackage{multirow}
\DeclareRobustCommand{\VAN}[3]{#2}
\let\VANthebibliography\thebibliography
\def\thebibliography{\DeclareRobustCommand{\VAN}[3]{##3}\VANthebibliography}

\usepackage{graphicx}	%
\usepackage{amsmath}	%

\usepackage{natbib} %
\usepackage{epsfig,graphicx}
\usepackage{epstopdf}
\usepackage{subcaption}
\usepackage{float}
\usepackage[usenames,dvipsnames]{xcolor}
\usepackage{lastpage}
\usepackage{bm}

\DeclareGraphicsExtensions{.pdf,.png,.jpg,.eps}
\graphicspath{{figures/},{figs_deltagp1/},{figs_deltagp1/figs_updated/}}%

\usepackage[dvipsnames]{xcolor}
\usepackage{comment}
\usepackage[normalem]{ulem}

\newcommand{\mean}[1]{\ensuremath{\langle #1 \rangle}}

\newcommand{\wgp}{\ensuremath{w_{g+}}}

\newcommand{\mpch}{\ensuremath{h^{-1}\text{Mpc}}}

\newcommand{\DS}{\ensuremath{\Delta\Sigma}}

\title[Model-Independent IA]{Model-Independent Assessments of Tidal Alignment and Density-Weighting Dependence in Galaxy Intrinsic Alignments}

\author[Rogozenski et al.]{Paul Rogozenski$^{1}$\thanks{\tt progozen@andrew.cmu.edu},
Sukhdeep Singh$^{1,2}$,
Jonathan Blazek$^{3}$,    
Rachel Mandelbaum$^{1}$,
Hung-Jin Huang$^{4}$, \newauthor
Ananth Tenneti$^{1}$  \\
$^{1}$McWilliams Center for Cosmology and Astrophysics, Department of Physics, Carnegie Mellon University, Pittsburgh, PA 15213, USA\\
$^2$ Berkeley Center for Cosmological Physics, University of California, Berkeley, CA 94720, USA\\
$^3$ Department of Physics, Northeastern University, Boston, MA, 02115, USA\\
$^4$Steward Observatory/Department of Astronomy, University of Arizona, 933 North Cherry Avenue, Tucson, AZ 85721, USA\\
}

\date{Accepted XXX. Received YYY; in original form ZZZ}

\pubyear{2026}

\begin{document}
\label{firstpage}
\pagerange{\pageref{firstpage}--\pageref{lastpage}}
\maketitle

\begin{abstract}
  Intrinsic alignments (IA) of galaxies are an important source of systematic uncertainty in weak gravitational lensing measurements and also serve as a test for models of galaxy formation and evolution. These galaxy alignments are usually modeled assuming that the IA are sourced by the tidal fields around the galaxies. However, galaxy properties, like their proximity to other galaxies and their mass, provide alignment contributions beyond the tidal field. %
  In this work, we perform novel model-agnostic tests to determine the impact of density-weighting on the IA of galaxies using data from the IllustrisTNG cosmological hydrodynamical simulation and SDSS-III (BOSS). We show that the primary effect of density-weighting is an increase of the alignment amplitude at large scales, %
  whereas the small-scale impact is sample-dependent. We also assess the tidal alignment paradigm for IllustrisTNG galaxies using two-point correlation functions of galaxy shapes, tidal fields, and the sampled background tidal field of the simulation box. We find, given the precision of our measurements, that central galaxies align linearly to the underlying (non-linear) tidal field well into the 1-halo regime.
\end{abstract}

\section{Introduction}
	In the next 5--10 years, several %
    cosmological surveys aim to utilize weak gravitational lensing, 
	the deflection of light  due to the presence of matter, to make precise measurements of 
	structure growth and its evolution over cosmological time scales. Among these surveys are the Vera C. Rubin Legacy Survey of Space and Time \citep[LSST,][]{2019ApJ...873..111I}, the Roman Space Telescope High-Latitude Wide-Area Survey \citep[HLWAS,][]{2015arXiv150303757S, 2025arXiv250510574O}, and the European Space Agency's Euclid mission \citep[][]{2025A&A...697A...1E}.  %
    These upcoming measurements will allow us to test aspects of the cosmological model related to dark matter and dark energy \citep[see][for a review]{Kilbinger2015}, as well as the galaxy-halo connection \citep[see][for a review]{Wechsler_2018}, with unprecedented precision.

	One important source of systematic uncertainty in weak lensing measurements   is the intrinsic alignment
	of galaxies (IA hereafter), which contaminates
	the weak lensing signal and hence can bias the inferred cosmological parameters from weak lensing measurements if not properly modeled 
	\citep[see][for reviews]{Troxel2015,Joachimi2015ia, 2024OJAp....7E..14L, Chisari_2025}. %
    Formally, IA is defined as the coherent alignment of galaxies with nearby galaxies and the large-scale structure due to physical causes other than lensing, and this effect can manifest itself as a false weak lensing signal.  Environmental effects, particularly the effects of the local tidal field, are understood to be the dominant sources of IA \citep[e.g.,][]{Croft_2000,Catelan2001}. %
    Based on this understanding, \cite{Catelan2001} and \cite{Hirata2004} developed linear models for IA sourced by the 
	tidal fields, assuming that the tidal fields at the time of galaxy formation set the IA. Sufficient smoothing is applied to the tidal field to exclude galactic-scale fluctuations, although selecting a smoothing scale is not an obvious choice \citep[as investigated in e.g.,][]{Blazek2011, Chisari_2013,zjupa-ia, 2023MNRAS.523.5899D}.
 Later, \cite{Bridle2007} replaced the linear power spectrum with the non-linear power spectrum, resulting in the non-linear alignment (NLA) model. Terms beyond the first-order alignment of galaxies to the underlying tidal field especially become important towards smaller scales where density weighting (the fact that shape measurements can only be measured at galaxy locations) and tidal torquing (dependencies proportional to the quadratic tidal field, usually in angular momentum-supported systems) become important. 
 Further evolution of this model family led to the Tidal Alignment and Tidal Torquing (TATT) model,   %
 which accounts for effects proportional to density-weighting and a quadratic term in the measurements \citep{Blazek2015,tatt}.
 There are other modeling approaches to IA as well, such as machine learning-based models \citep[e.g.,][]{graphgan}, effective field theory-based \citep[e.g.,][]{eft-ia, 2023JCAP...10..005B}, and halo models for small scales \citep[e.g.,][]{fortuna,nick-ia}. 

Two-point correlations are often leveraged to understand galaxy IA by correlating galaxy positions with galaxy shapes to measure the alignment strength of galaxy populations at a given separation scale. 
Galaxy intrinsic alignments are ``directly'' measured in data via two-point correlation functions \citep[see e.g.,][]{Joachimi2011, Singh2015, Singh2016ia, kids-gama-ia, des-y3-ia, 2025A&A...699A.252G, 2026MNRAS.545f1630N} and self-calibration techniques \citep[e.g.,][]{Yao2019,2026arXiv260110314P} with careful redshift selection and are likewise measured in hydrodynamical simulations \citep[e.g.,][]{ananth-subgrid,tenneti-ia,samuroff-2020, bulge-disc, 2026A&A...711A.235H}. The simplest approach to understanding the underlying mechanisms of IA is to fit a model (commonly NLA or TATT) to the standard two-point projected position-shape correlations. Cosmological inferences implement IA models to account for their systematic uncertainty in the cosmic shear signal, ``indirectly'' constraining IA properties \citep[e.g.,][]{hsc-y3-li,2025A&A...703A.158W, desy6_cosmic_shear}%

However, these data have not yet shown a consistent analytic model preference. %
Recent indirect measurements of IA from cosmological weak lensing constraints show a mild preference for modeled density-weighted contributions to have the opposite sign as expected \citep{desy6_cosmic_shear} and continue to show no preference across a range of different models, including additional scale-dependent and mass-dependent terms \citep{2025A&A...703A.158W}. %
 Observational data from the real universe provide insights into genuine observable galaxy properties and their alignment, but they often suffer from limited SNR, limited redshift information, and a lack of information on the underlying dark matter distribution.  Fitting additional parameters to better model the non-linear clustering of galaxies or baryonic effects suffers from the dimensionality of fitting additional parameters.  %
 These approaches are generally model-dependent and provide limited insight into the individual underlying physical mechanisms that drive the galaxy alignment amplitude.

 Some of the modelled contributions of the NLA and TATT models have been examined in hydrodynamical simulations, where constraints on the terms of IA models can be obtained via direct one-point correlations of galaxy shapes with tidal fields \citep[e.g.,][]{samuroff-2020,zjupa-ia, 2023MNRAS.523.5899D}. %
 These one-point approaches are informative on how the distributions of galaxy shapes and the underlying tidal fields are related, however they can be difficult to interpret in the context of the two-point (and higher) correlations directly relevant for lensing measurements.
 Calculating these correlations in hydrodynamical simulations provides a great amount of `truth' information for model-agnostic analyses \citep[insofar as choices of sub-grid physics and amplitudes are model independent, but see e.g.,][for discussion on how these choices affect IA]{ananth-subgrid, 10.1093/mnras/staa045,10.1093/mnras/stag229}.

The first goal of this paper is to present a method for direct IA constraints in hydrodynamical simulations %
that is not as model-dependent as fits to NLA or TATT. %
Because we can directly measure the tidal field at galaxy positions, the density-shape and density-tidal field correlations experience similar contributions due to small-scale effects (non-linear galaxy clustering, baryonic physics, etc.), enabling a clean model-independent probe of how 
galaxy shape alignments respond 
to the local tidal field. 
Hence, this becomes our empirical model that already includes non-linear effects, to be used in lieu of perturbative analytic models \citep[as in e.g.,][]{2026MNRAS.545f1630N}.

 The second goal of this work is to explore the density-weighting effects. %
 As was shown by \cite{2014AA...561A..53V,Blazek2015}, density weighting effects
	can naturally lead to some dependence on galaxy bias in the measured IA signal and has been explored by e.g., looking at the halo mass relationship to the large-scale alignment amplitude \citep[][]{2017MNRAS.468.4502V} or jointly fitting the linear galaxy bias with an IA amplitude for different luminosity and density cuts \citep{kids-gama-ia, fortuna}. Isolating the effect of density weighting on IA can help elucidate one of the most interesting, yet not fully explained, observations: the evolution of the strength of IA with the galaxy bias or halo mass  \citep{Joachimi2011,Singh2015}, where the galaxies that are brighter or are in more overdense regions tend to show stronger alignments. 

    In this work, we explore assumptions in intrinsic alignment modeling using model-agnostic two-point correlations.  We first conduct measurements to support our two goals in cosmological hydrodynamical simulations \citep[IllustrisTNG;][]{tng-publicdata} by computing position-shape and position-tidal field correlations for different mass and morphology samples. We demonstrate how ratios of these correlation functions enable us to explore the effect of density-weighting and galaxy clustering across multiple samples and assess the alignment mechanisms across the accessible scales of the simulation volume. We also demonstrate how to explore model assumptions through real observations \citep[SDSS-III BOSS Data Release 12;][]{sdss-dr12} from the assembly of density-weighted shear maps. Thus, our novel approaches achieve model-independent tests of the tidal alignment assumption with tidal fields containing realistic/real galaxy properties, galaxy clustering, and small-scale effects. %

We begin in \S~\ref{theory} with the description of the theoretical background and an introduction on the two-point correlations used in this work. Next, in \S~\ref{data} the simulation suite and the observational data are described. In \S~\ref{estims} we describe the various estimators that were used to measure IA. 
We present and interpret the measured two-point correlation functions from simulations in \S~\ref{tng-2pt} and from observational data in \S~\ref{boss-2pt}, %
showing the results of our IA model assumption tests. 
Finally, in \S~\ref{conc} we discuss and conclude our paper.  
For the analyses of real data and theoretical predictions, we use the Planck-2015 cosmology \citep{Planck2015cosmo}, the fiducial cosmology of IllustrisTNG, %
 with distance measurements in comoving \mpch.

	\section{Theoretical background and Two-Point Correlations}\label{theory} 
		We discuss the basic formalism for describing the physics determining the galaxy shear (IA and lensing), the effects of the density weighting \citep[closely following the 
		discussion in][]{Blazek2015}, and the basic formalism of two-point functions used to measure galaxy-shape alignments along with their covariances. 
  
    	\subsection{Galaxy-Shear Cross-Correlations and Density Weighting}\label{sec:shear}
			The shearing of galaxy shapes, written as $\gamma = \gamma^{\rm I} + \gamma^{\rm G}$, sourced by both IA and weak gravitational lensing, respectively,  is assumed to be linearly proportional to %
            the gravitational tidal field to leading order. These quantities are implicitly functions of configuration-space position, which we omit for compactness; quantities with a tilde denote Fourier-space analogs. The difference between the two terms is that the IA is sourced by tidal fields within the local environment of the galaxies,  while the gravitational lensing is sourced by the tidal effects of the matter between the galaxy and the observer \citep[for a review, see for example][]{Troxel2015}. 
                    
				Within the tidal alignment model, the IA shear is sensitive to the tidal field sourced by the local environment of the galaxies for which shear is measured.
				By convention, the IA shear is written as
            	\begin{align}\label{eq:gammA_Ia}
            		\gamma^{\rm I}(z)=\left(\gamma_+^{\rm I}(z),\gamma_\times^{\rm I}(z)\right)=
                    			-A_{\rm I}\frac{C_1}{4\pi G D(z)}\left(\partial_x^2-\partial_y^2,2\partial_x\partial_y\right)\phi (z) \\
					\tilde\gamma^{\rm I}(z)=\left(\tilde\gamma_+^{\rm I}(z),\tilde\gamma_\times^{\rm I}(z)\right)=
                    			-A_{\rm I}\frac{C_1\bar{\rho}_m}{D(z)}\left(\frac{k_x^2-k_y^2}{k^2},\frac{2k_xk_y}{k^2}\right)\tilde \delta_m (z)
            	\end{align}
                where the first equation is in configuration space and the second is in Fourier space. We also used the Poisson equation, $\rho_m(z)=\nabla^{2}\phi(z)$ to convert from the potential to 
               the matter overdensity. $A_{\rm I}$ is the unknown alignment amplitude to be constrained using data, and $D(z)$ is the growth factor at redshift $z$. We have implicitly substituted $\phi_p=\phi(z)/D(z)$ into the above equations, where $\phi_p$ denotes the potential at the time of the formation of the galaxy and when its initial intrinsic alignment is set. %
               $C_1$ is a fixed constant (usually together with $\rho_\text{crit}$) representing the expected order of magnitude
of the tidal response of galaxy shapes \citep[originally from shape variance measurements;][]{Brown_2002} and $A_{\rm I}$ can be a general sample-dependent function of redshift, mass, etc. \citep[as discussed in e.g.,][]{Blazek2015}. %
The `$+$' component denotes radial/tangential alignment of the field, with a positive/negative $A_{\rm I}$ denoting a radial/tangential alignment in our sign convention. The `$\times$' component is the cross-component alignment of the field (i.e., $45^\circ$ offset from the `$+$' axes). %

				The weak gravitational lensing shear induced by a lens at redshift $z_l$ on a source at $z_s$ is given as 
            \begin{align}
            \label{eq:gamma_lens}
            		\gamma^{\rm G} ( z_l, z_s) &=\left(\gamma_+^{\rm G}( z_l, z_s),\gamma_\times^{\rm G}( z_l, z_s)\right) \nonumber\\&=
                    	\frac{1}{4\pi G \Sigma_c( z_l, z_s) }(\partial_x^2-\partial_y^2,2\partial_x\partial_y)\phi_{l} (z_l)
            	\end{align}
 with $\phi_{l}$ denoting the lens potential at redshift $z_l$. The comoving distances to the lens, the source and between the lens and source are written as $\chi_l$, $\chi_s$ and $\chi_{ls}$, respectively, and are functions of the lens and source redshifts. The critical surface mass density, $\Sigma_c$, is defined as
          \begin{align} \label{sigma_crit}
            \Sigma_c( z_l, z_s) = \frac{c^2}{4 \pi G}   \frac{ \chi_s }{ \chi_l   \chi_{ls}  }.
        \end{align}
 This quantity describes the efficiency of gravitational lensing, set by the distances between the observer, lens, and source; the weak gravitational-lensing regime holds when the surface mass density of the lens plane is much smaller than this value. We also define quantities $\delta_g$ and $\delta_s$ as the galaxy number overdensity fields of the lens and source samples, respectively. To linear order, this is proportional to the matter density contrast via the sample's galaxy bias, $b_g$, as $\delta_g = b_g \delta_m$. We will generally use a subscript of $\delta$ to denote different galaxy samples (either a fixed density tracer sample $g$ or a shape sample, $s$), but $\delta_m$ denotes the matter density contrast.

We estimate the shear by measuring shapes at galaxy locations; hence the observed shear field is weighted by the density of the source galaxies. The density-weighted shear field, $\gamma^w$, is then
\begin{equation}
    \gamma^w(z_l, z_s)=\left(1+\delta_s(z_s)\right)\gamma(z_l, z_s). %
\end{equation}

The most common estimator for galaxy-shear cross-correlations for both lensing and IA measurements involves cross-correlating the galaxy positions
            with the shapes of the galaxy sample tracing the shear (hereafter shape sample)
            and then normalizing by the expected number of galaxy pairs using either the random-random correlations or
            random-source correlations \citep[e.g.,][]{xi-gp}:
  \begin{equation}
  \label{eq:xigp}
      \xi_{g+} =\mean{\widehat\delta_g\widehat \gamma_+^w}=\frac{S_+D-S_+R}{R_SR_D}.
  \end{equation}
For the observed density tracer sample, $\widehat\delta_g$, we can compute the cross-correlations between the shear field as 
\begin{align}
    \label{eq:gamma_expansion}
    \nonumber \mean{\widehat\delta_g(z_l)\gamma_+^w(z_l,z_s)}&=\mean{\widehat\delta_g(z_l){\gamma_+}(z_l,z_s)}+\mean{\widehat\delta_g(z_l)\delta_s(z_s){\gamma_+}(z_l,z_s)}\\ \nonumber
    &=\mean{\widehat\delta_g(z_l){\gamma_+^{\rm I}}(z_s)}+\mean{\widehat\delta_g(z_l)\delta_s(z_s){\gamma_+^{\rm I}}(z_s)} \\ &+ \mean{\widehat\delta_g(z_l){\gamma_+^{\rm G}}(z_l,z_s)}+\mean{\widehat\delta_g(z_l)\delta_s(z_s){\gamma_+^{\rm G}}(z_l,z_s)},
\end{align}
where we make the redshift dependencies of the relevant quantities explicit.

The density weighting terms arise because IA are a local phenomenon and are relevant only when density and shape tracer galaxies are close along the line of sight and, 
            thus, correlated. In this limit, $z_s \approx z_l$ and
              \begin{equation}
              \label{eq:wgp_approx}
                  \mean{\widehat\delta_g (z_l) \gamma_+^w (z_l)}\approx\mean{\widehat\delta_g(z_l){\gamma_+^{\rm I}(z_l)}} + \mean{\widehat\delta_g(z_l)\delta_s(z_l){\gamma_+^{\rm I}(z_l)}}.
              \end{equation}
              Note that $\delta_s$ and $\gamma_+^{\rm I}$ are fields at the same physical position and may not be expanded as a typical bispectrum \citep[see App. A of][for an analytic expansion in Fourier space]{Blazek2015}.
              For weak lensing measurements, however, we select lens and source galaxy pairs that are widely separated along the line of sight.  For these circumstances, the density weighting term is
              small and thus we get
              \begin{equation}
              \label{eq:dsig_approx}
                  \mean{\widehat\delta_g (z_l) \gamma_+^w (z_l,z_s)}\approx\mean{\widehat\delta_g(z_l){\gamma_+^{\rm G}(z_l, z_s)}}
              \end{equation}
Equations~\eqref{eq:wgp_approx} and~\eqref{eq:dsig_approx} projected along the line of sight can be interpreted as either the position-shape cross-correlation $w_{g+}$ or the excess surface density $\Delta \Sigma$ when appropriately weighted by $\Sigma_c$ (Eqn.~\ref{sigma_crit}), respectively, for some projected separation length $r_p$. 

In this configuration, both correlations share the same lens sample and respond to the same lens potential.
Equations~\eqref{eq:gammA_Ia} and~\eqref{eq:gamma_lens} can be substituted at linear order to give the relation 
\begin{equation}
    {\mean{\widehat\delta_g(z_l){(1+\delta_s(z_l))\gamma_+^{\rm I}(z_l)}}} = - A_{\rm I} \frac{C_1}{D(z_l)} \Sigma_c^{}(z_l, z_s){\mean{\widehat\delta_g(z_l){\gamma_+^{\rm G}(z_l, z_s)}}},
\end{equation}

Integrating both sides of the equation over the line-of-sight positions of the source sample for some projected separation $r_p$ (and recalling each side of the equation has non-zero weights depending on the line-of-sight separation of the lens-shape pair), we arrive at the relationship
\begin{equation}\label{eq:gamma_relation}
    \wgp(r_p, z_l) = -A_{\rm I} \frac{C_1}{D(z_l)}\DS(r_p, z_l).
\end{equation}
We require $z_b > z_a+\delta z$ for the $\DS$ estimator, and that $a,b$ are close pairs that exist within $\pm \Pi_{\rm max}$ along the line of sight. 
              We obtain \wgp by integrating $\xi_{g+}$ over the line-of-sight direction:
              \begin{align}
              \label{eqn:wgp}
              	w_{g+}(r_p)=\int_{-\Pi_\text{max}}^{\Pi_\text{max}} d\Pi\,  w_g(r_p,\Pi)  w_+(r_p,\Pi) \,\xi_{g+}(r_p,\Pi),
              \end{align}
writing generally to allow additional sample-dependent weight functions along the line-of-sight, $w_X(r_p,\Pi)$. 
A sufficiently large $\Pi_{\rm max}$ is typically used to avoid the impact of redshift space distortions. 
\wgp\ is measured in narrow $\Pi$ bins (compared to the much broader line-of-sight separations used for $\Delta\Sigma$) 
       and the IA contamination in galaxy-galaxy lensing observables
       are only consistent when the line-of-sight window is uniform \citep{probing-gravity}. 
       The non-uniform window has a smaller effect on \wgp\ as it is removed after division with the $RR$ pair counts, which   is not the case for the $\Delta\Sigma$ estimator. Hence, the weights in the two estimators may also include implicit $\Pi$ dependence.      
        We do not utilize a random catalog when measuring the correlations within the simulation box. Instead, we use the analytic expectation of the $RR$ term, as in \cite{shakir}, since we do not need actual randoms with an ideal periodic box geometry. %

Given the cosmological principle of homogeneity and isotropy, the correlation functions depend only on the magnitude of the separation vector in configuration space. However, in real data, the projection of galaxy shapes from 3D to 2D on the sky introduces additional anisotropy in the observables. Galaxy positions are also estimated using redshifts, which are contaminated by coherent distortions due to the peculiar velocity of galaxies. While our real data analyses are done in redshift space, our simulation analyses are done in configuration space directly \citep[see][for more details and description]{Singh2016ia, shakir, 2024MNRAS.534.3540L}.  

Alternatively, one can expand $\xi_{g+}$ in multipoles \citep[][]{shakir} or assess galaxy multiplet alignments \citep[][]{2024MNRAS.534.3540L}, which extract more information from the 3D field and thus yield higher signal-to-noise ratios but is beyond the scope of this paper. %

        \begin{figure}
            \centering
            \includegraphics[width=1.0\linewidth]{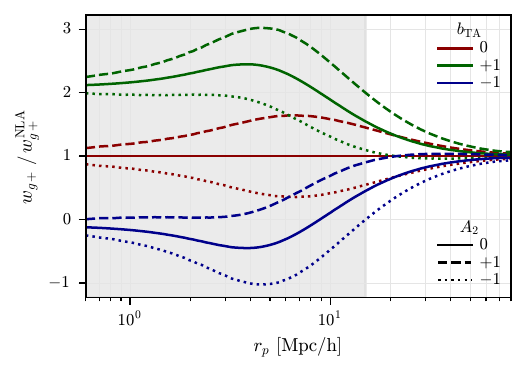}
            \caption{The response in \wgp\ statistics to beyond-linear model parameters relative to correlations using the NLA model. Dashed, solid, and dotted lines correspond to realizations model parameters $A_2=1,0,-1$, respectively, while green, red, and plot lines correspond to $b_{\rm TA} = -1,0,1$, respectively, for a constant $b_g = A_{\rm I} = 1.5$. While the various model realizations converge at very large scales, divergences are seen as we approach the scales analyzed in the TNG100-1 simulation analyzed in this work (filled in gray).}
            \label{fig:tatt_ratios}
        \end{figure}

\subsection{NLA and TATT } 
        Here we describe how the TATT model incorporates the density weighting effect compared to the NLA model. Both models assume that galaxy alignments are set by the tidal tensor of the large-scale structure:
        \begin{equation}
            \label{eqn::general_tidal_field}
            \tilde s_{ij} (\bm{k},z) = \Big ( \frac{k_i k_j}{k^2} - \frac{1}{3} \delta_{ij} \Big) \tilde\delta_{\rm m}(k,z)
        \end{equation}

         Using a perturbative approach, the intrinsic galaxy shear can be written in terms of the tidal tensor as follows (up to quadratic order):
        \begin{equation}
            \tilde \gamma_{ij} \approx\bar{A}_1(z) \tilde s_{ij} + \bar{A}_{1\delta}(z) \tilde \delta_{\rm m} \tilde s_{ij} + \bar{A}_2 (z) (\tilde s_{ik} \tilde s_{kj} -\frac{1}{3} \delta_{ij} \tilde s^2).
            \label{eqn:tatt}
        \end{equation}
        This equation is written in 3D, using the 3D tidal tensor, as a function of $k$ at a given redshift, $z$. One can project this onto 2D to exclude the unobservable line-of-sight shear and obtain the standard $(\gamma_+^{\rm I},\gamma_\times^{\rm I})$.  For the explicit form and details refer to \cite{tatt} \S II.C. Note that in the above formulae, $\delta_{ij}$ refers to the Kronecker delta and the other instances of $\delta$ denote overdensity. 

        In the above formula, we have indicated that the alignment amplitude parameters have the model freedom to be functions of redshift (e.g., $\bar{A}_1(z) =  A_1\frac{C_1\bar{\rho}_m}{D(z)}$, $\bar{A}_{1\delta}(z) =  A_{1\delta}\frac{C_1\bar{\rho}_m}{D(z)}$, and $\bar{A}_2(z) =  5A_2\frac{C_1\bar{\rho}_m}{D(z)^2}$). Modern analyses also include some power function of redshift, luminosity, or halo mass dependence to the model \citep[as in e.g.,][]{2025A&A...703A.158W, 2026PhRvD.114f3506S}.
        Here $A_1$, $A_{1\delta}$, and  $A_2$  are the free amplitude parameters of the model, which capture the strength of the alignment due to the corresponding overdensity (scalar) and tidal (3x3 matrix) fields. 
        $A_1$ describes the tidal alignment strength, equivalent to the amplitude prefactor of Eqn.~\eqref{eq:gammA_Ia} ($A_1 = A_{\rm I}$). %
        $A_2$ describes the tidal torquing strength and the $A_{1\delta}$ term describes the density-weighting arising from the fact that galaxy shapes are measured only where there is a galaxy; these amplitudes also encapsulate any other effects with the same density and tidal field dependencies. %
        The TATT model uses all of the parameters in Eqn.~\eqref{eqn:tatt}, whereas the NLA model uses only $A_1$ with $A_2 = A_{1\delta} = 0$ \citep{tatt}. In some papers, $A_{1\delta}$  is parametrized as $A_{1\delta} = b_\text{TA} A_{1}$, in which case $b_\text{TA}$ becomes the TATT fitting parameter for density weighting rather than $A_{1\delta}$. 
        In Figure \ref{fig:tatt_ratios}, we show an example of how the individual analytic components of the TATT model affect the alignment amplitude relative to the NLA model. We do not explicitly fit these models in this work; instead, we take a model-agnostic approach to studying them, as detailed in section~\ref{estims}.

\section{SDSS Data and IllustrisTNG Simulations}\label{data}
 \subsection{Simulated data: IllustrisTNG}\label{simulation}

\begin{figure*}
    \centering
    \includegraphics[width=1\linewidth]{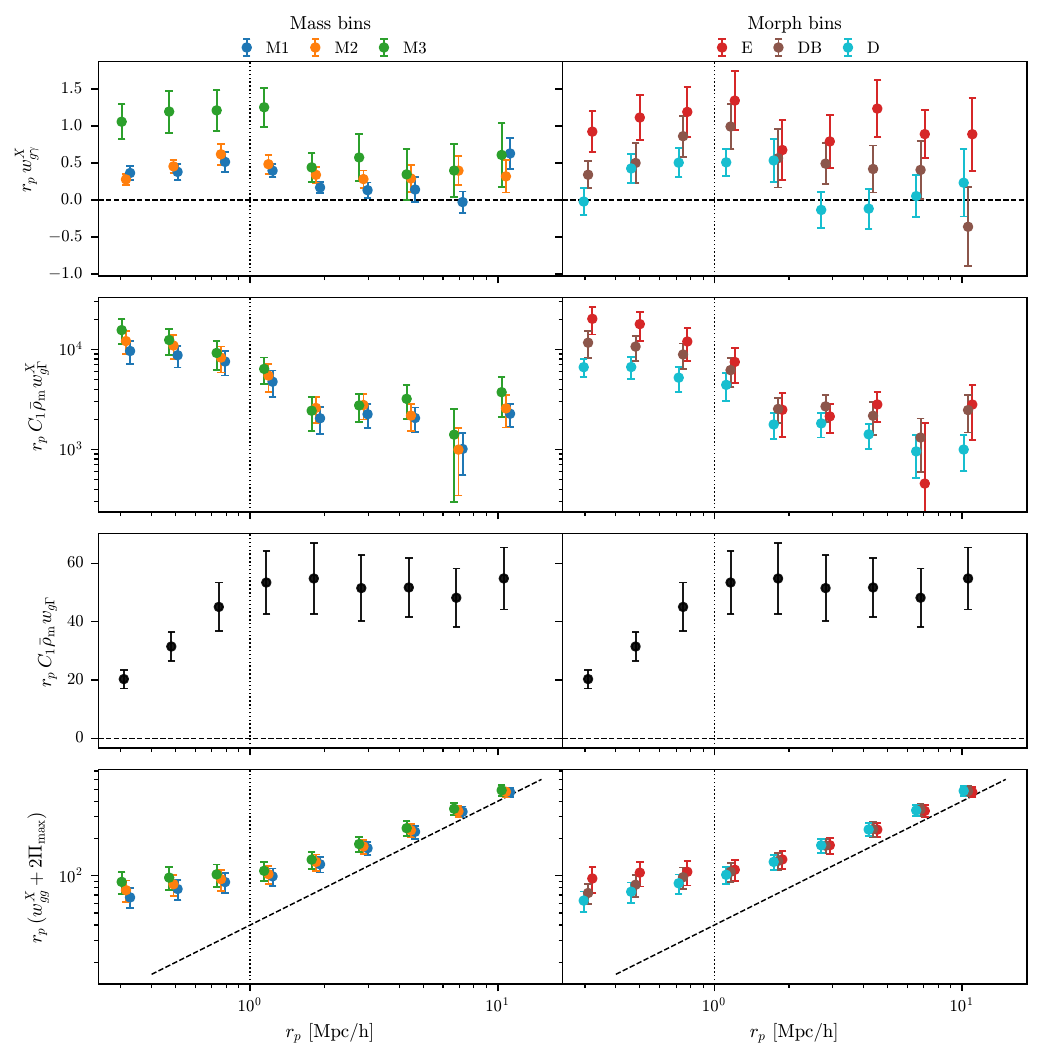}
    \caption{The measured correlation functions in TNG100-1 relating the density tracer galaxy sample to the alignment of the galaxy shape (first row) and the underlying galaxy tidal field (second row) for the galaxy sub-samples considered in this work (see Section~\ref{simulation}). We also show the density-shape correlation of the background tidal field (third row) and the galaxy clustering correlation between the total density sample and the shape galaxy's position (fourth row), with the large-scale clustering expectation value $2~r_p\Pi_{\rm max}$ as the diagonal dashed line. Correlations involving tidal field measurements are weighted by $C_1 \bar{\rho}_\mathrm{m}$. 
    We depict only the points that are used in our analysis (see Section~\ref{tng-2pt}), where data points are horizontally shifted for readability. %
    We fit ratios of these data for either all plotted points or for $r_p\geq 1 \mathrm{Mpc}/h$, denoted by the vertical dotted line. 
    }
    \label{fig:wgp_data}
\end{figure*}
 
In this work, we have used the IllustrisTNG simulation suite, namely the highest-resolution version TNG100-1
\citep[for more information, please refer to][]{ tng-bimodal,pillepich2018illustristng, Springel2017illustristng, Naiman2018illustristng, Marinacci2017illustristng,tng-publicdata}.
TNG100-1 is a large-volume gravo-magnetohydrodynamical simulation with a box side length of 75 $h^{-1}$Mpc. The simulation was run using the moving-mesh code Arepo \citep{arepo} and  has $2 \times 1820^3$ resolution elements with a gravitational softening length of 0.7 $h^{-1}$kpc 
for dark matter and star particles. 
The dark matter particle mass is $7.46\times 10^6  M_\odot$ and star particle masses are variable. The physical models and processes for galaxy formation and evolution include radiative gas cooling and heating; star formation in the ISM; stellar evolution with metal enrichment from supernovae; stellar, AGN and blackhole feedback; formation and accretion of supermassive blackholes \citep{tng-methods,tng-agn}.
The halos within the simulation are identified through the friends-of-friends (FoF) 
method \citep{fof}, and then the subhalos  
are identified using the SUBFIND algorithm \citep{subfind}. 

            \begin{table}
              \centering
              \caption{Galaxy central fraction, $f_{\rm cen}$, and total number of galaxies, $N_{\rm tot}$ for each galaxy sub-sample.}
              \label{tab:cen_frac}
              \begin{tabular}{ l|c|c|c|c|c|c| }
                \hline
                Sample & M1 & M2 & M3 & E & DB & D\\
                \hline
                $f_{\rm cen}$ & 0.543 & 0.565  & 0.641  & 0.615  & 0.561 & 0.649 \\
                $N_{\rm tot}$ & 9208 & 9558 & 2870 & 2152 & 2152 & 2152 \\
                \hline
              \end{tabular}
            \end{table}

The simulation suite includes 100 snapshots at different redshifts; we use the latest snapshot at $z=0$ for our analysis. To study the impact that density-weighting has on the relationship between galaxy properties (morphology, mass) and galaxy shape alignments to the projected tidal field, we split %
$\wgp$ measurements into different subdivisions. We consider three stellar mass bins when measuring the density-weighted tidal field and shapes of galaxies, defined as $M_1 = 10^{9-9.5} h^{-1}M_{\odot}$, 
			$M_2 = 10^{9.5-10.5} h^{-1}M_{\odot}$ and $ M_3 \geq 10^{10.5} h^{-1}M_{\odot}$. We also analyze morphologically split samples with stellar mass $\geq 10^{10}  h^{-1} M_\odot$, with indicators  `E' for elliptical galaxies, `DB' for two-component disc+bulge galaxies, and `D'  for pure disc galaxies. The morphology samples, as defined in \cite{bulge-disc}, are determined kinematically by evaluating the angular momenta of stellar particles and determining the fraction of stars in the disc component of the galaxy, $f_\text{disc}$, and are mass-weighted to match the disc sample. The total morphology sample was split equally into thirds defined by $f_\text{disc}$, where the lowest, middle, and highest values of $f_\text{disc}$ determined the E, DB, and D samples, respectively. Kinematically-defined morphological samples are important to investigate, as galaxy alignment mechanisms are expected to depend on the galaxy's angular momentum \citep{Hirata2004} %
            For a given measurement of $\wgp$, we modify the subscript to denote the measured field, either the galaxy-shape correlation $g\gamma$ or the galaxy-tidal field correlation $g\Gamma$, and a superscript, $X$ or $Y$, to denote any galaxy property selection. For example, the $\wgp$ statistics of the galaxies' underlying tidal field for the highest-mass bin will be written as $w_{g\Gamma}^{M3}$ %
            We also decompose these samples into central and satellite sub-samples, denoted by an additional (cen) or (sat) superscript, which helps determine galaxy alignment properties based on the galaxy's location in the halo. The fraction of central galaxies in each subsample, $f_{\rm cen}$, is provided in Table ~\ref{tab:cen_frac}. More information about the morphological classification and justifications for the lower mass limits can be found in \cite{decomp} and \cite{bulge-disc}, respectively. 
         
            We perform two sets of tidal field correlations, one where the smoothed tidal field (Eqn.~\eqref{eqn:smoothed-tidal-field}) is measured at galaxy locations and another where it is measured at random locations throughout the box. Measuring the $w_{g\Gamma}$ correlation of the tidal field at random points to the total galaxy sample measures the non-density-weighted, or background, tidal field correlation, with which we can make quantitative statements about dependencies of the density-weighted tidal field on the alignment of galaxies. The unweighted background correlation will have no superscript and is denoted simply as $w_{g\Gamma}$. Our fiducial analysis considers a Gaussian smoothing scale of $r_\mathrm{s} = 0.1 \rm{Mpc}/h$ (see section \ref{tidal-shear} for more details).%
            We plot these correlations in Figure \ref{fig:wgp_data}.

		\subsection{Real data: SDSS} \label{sec:data_sdss}

The Sloan Digital Sky Survey  \citep[SDSS;][]{2000AJ....120.1579Y} consisted of   multi-band drift-scanning imaging \citep{1996AJ....111.1748F, 1998AJ....116.3040G, 2001AJ....122.2129H, 2002AJ....123.2121S,2004AN....325..583I} and follow-up spectroscopy \citep{2001AJ....122.2267E,2002AJ....123.2945R,2002AJ....124.1810S} using the SDSS Telescope \citep{Gunn2006}.  The survey covered roughly $\pi$ steradians
of the sky. %
  
The BOSS survey made spectroscopic measurements of selected galaxies from the imaging \citep{Ahn:2012,Dawson:2013,Smee:2013,Bolton:2012}. 
In this paper, we used SDSS-III BOSS Data Release 12 (DR12) %
LOWZ galaxies \citep{sdss-dr12} in the redshift range $0.16<z<0.36$.%
The LOWZ sample is approximately volume-limited in the redshift range $0.16<z<0.36$ and consists of Luminous Red Galaxies (LRGs), with a number density of $\bar{n}\sim 3\times10^{-4}h^3\text{Mpc}^{-3}$ \citep{Manera2015}. %
  We obtained the intrinsic alignment measurements by combining galaxy shape measurements from \cite{Reyes2012} with spectroscopic redshifts from BOSS. After applying redshift cuts and extinction cuts, the LOWZ sample consists of 225181 galaxies, of which there are usable shape measurements for 181933 galaxies. For derived quantities  such as rest-frame colours and luminosities, please see \citet{Singh2015,Singh2016ia} for more details. %

  Finally, when computing $\Delta\Sigma$ with BOSS galaxies, redshift estimates for the fainter sources are needed. For this purpose, the maximum-likelihood estimates of photometric redshifts were obtained based on the five-band photometry from the 
Zurich Extragalactic Bayesian Redshift Analyzer \citep{zebra}.  For more details on the photometric redshifts please refer to \cite{Nakajima2012};  further steps to calculate the signals and remove biases due to use of photometric redshifts follow the approach of \cite{probing-gravity}. %

 \subsection{Jackknife Covariance Estimates}
 
We estimated the covariance matrices for the SDSS measurements using the jackknife method.   
First, the survey is divided into  $N_\text{jk}=100$ equal-area regions. %
 Second, the measurements are re-done
 $N_\mathrm{jk}$ times, while dropping one jackknife region during each measurement, such that each time the measurements include galaxies from $N_\mathrm{jk} -1 $ regions. Given a data vector $\psi_i \in (\xi, w, \Delta\Sigma)$ -- with $i$ indexing the $r_p$ bins in our measurement  -- the covariance matrix is computed as
\begin{equation}
   C^\mathrm{jk}_{ij}  = \frac{N_\mathrm{jk}-1}{N_\mathrm{jk}} \sum^{N_\mathrm{jk}}_{m=1} \big( \psi^{(m)}_i - \overline{\psi}_i \big) \big( \psi^{(m)}_j - \overline{\psi}_j \big)
   \label{eqn:cov}
\end{equation}
where $\overline{\psi}_i = \frac{1}{N_\mathrm{jk}} \sum^{N_\mathrm{jk}}_{m=1} \psi^{(m)}_i $,   with $m$  indexing the jackknife regions. 

We also compute the covariance matrices for the TNG100-1 simulation measurements via jackknife methods, dividing the simulation box into $N_\mathrm{jk} = 49$ equal-area regions projected along the line-of-sight. 
Given the length scales we analyze in the simulation, we opt to utilize a jackknife covariance estimate as opposed to a Gaussian analytic formulation \citep[e.g.,][]{Singh2017cov} to capture non-linear effects in the covariance. While jackknife covariances can create additional noise in the covariance estimate, they have been shown to be broadly consistent with analytical descriptions for a variety of two-point correlation observables \citep[e.g.,][]{Singh2017cov, 2022PhRvD.105h3528P, 2022MNRAS.514.1289M}.
For both the simulations and the real data, %
we follow the same procedure to calculate the jackknife covariance of ratios of quantities, substituting a ratio of correlation functions for $\psi_i$. For all covariances, we apply an additional factor that approximately debiases the estimated covariance matrix due to the chosen number of jackknife patches and binning of the data \citep{Hartlap2007}. %

    \section{Analysis Approach}\label{estims}
    In this section, we describe the various estimators used in our study. First, we describe the method to measure tidal shear from the simulations and the ratios we take to investigate the impact of density-weighting. We then describe the method to estimate the density-weighted contribution from real data.  
        
        \subsection{Tidal shear and shape estimation in simulations}\label{tidal-shear}
            
            To obtain the tidal field at a given point in a simulation, we use the ParticleMesh\footnote{\url{https://github.com/MP-Gadget/pmesh}} code to paint the simulation box. We use a fiducial mesh length of 1024 grid points. From this mesh, we obtain the real-space tidal field at a given point off the mesh by interpolation. We are interested in obtaining the smoothed tidal field at a given point to reduce artifacts from simulation discretization. We use a 3D Gaussian Filter in combination with the Fourier space-transformed tidal field for a characteristic length scale, $r_\mathrm{s}$, as
            \begin{equation}
            \label{eqn:smoothed-tidal-field}
                \tilde s_{ij} = \Big ( \frac{k_i k_j}{k^2} - \frac{1}{3} \delta_{ij} \Big) \tilde\delta_m(k)e^{-k^2 r_\mathrm{s}^2 / 2}
            \end{equation}
            
            We then inverse Fourier transform the tidal field back into configuration space and take the 2D slice 
            of the field in the $x$-$y$ plane to obtain the projected real-space tidal field $s_{ij}$. Our fiducial analysis uses a 0.1 Mpc/h smoothing scale, but we also consider smoothing scales of 0.5 and 1.0 Mpc/h to ensure the robustness of our results to the chosen smoothing scale at large separation scales (Appendix \ref{apx:smooth}).

\begin{figure}
                \centering
                \includegraphics[width=1\linewidth]{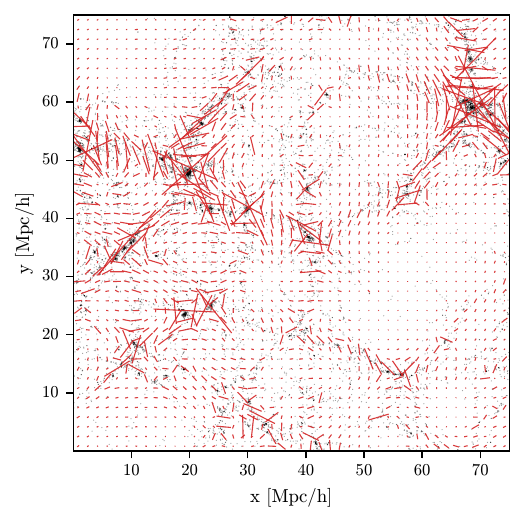}
                \caption{Visualization of projected ellipticities ($e_{\mathrm{tf}}$) of tidal fields in the TNG100-1 simulation. Average ellipticities are calculated over the z-range of 48.75-63.75 Mpc/$h$ (displayed as red sticks with a linearly scaled length), and galaxy locations are displayed by black dots. The $e_{\mathrm{tf}}$ amplitude grows with proximity to large-scale structure and tidal fields align along filaments of large-scale structure. %
                }
                \label{fig:sim_field}
            \end{figure}

           The correlation functions we obtain include the amplitude of the underlying shape/field as well as its radial orientation towards other objects. The shape amplitudes are bounded by the galaxy ellipticity, which is a positive number bounded from zero to one. In contrast, the amplitude of the tidal field is proportional to the amplitude of the matter density contrast and has no upper-bound for its value. The signal amplitude we observe is much greater for tidal fields at galaxy locations/overdense regions than for galaxy shape measurements, as seen in Figure \ref{fig:wgp_data}. 
           Although it is not an ellipticity by definition, we will often refer to a "tidal field ellipticity", which decomposes the real-space projection of Eqn.~\eqref{eqn:smoothed-tidal-field} into equivalent $e_{1,\mathrm{tf}}$ and $e_{2,\mathrm{tf}}$ components, where $e_{1,\mathrm{tf}} = s_{xx} - s_{yy}$, $e_{2,\mathrm{tf}} = 2 s_{xy}$ with $z$ as the line-of-sight direction. The amplitude of the tidal field ellipticity is then $|e_{\mathrm{tf}}| \equiv \sqrt{e_{1,\mathrm{tf}}^2 + e_{2,\mathrm{tf}}^2}$. We compute the projected tidal field both at galaxy locations and at randomly sampled points throughout the simulation box, enabling us to discern differences between the tidal field in which the galaxy resides and the background non-linear tidal field at a given separation scale. We visualize this definition of the projected tidal field in Figure~\ref{fig:sim_field}, where we plot the average projected ellipticity of the tidal field for a small slice of the TNG100-1 simulation box alongside the galaxies in that slice.

            Galaxy shape measurements are determined following the procedures of \citep{bulge-disc}, using the simple inertia tensor method of shape measurements, where we define ellipticities based on the decomposition of the projected inertia tensor as in \cite{2013MNRAS.431..477J}. %

 \subsection{Simulation ratio tests}

This study investigates the linear response of galaxy intrinsic shapes to the tidal field and assesses how density weighting affects the measured correlation functions. %
            To do so, we fit ratios of the measured correlation functions to a constant, $B$, and use the goodness of fit to assess whether a linear relationship is sufficient. Maintaining the same total density tracer sample for all analyses, we measure four such ratios for each galaxy subsample (summarized in Table \ref{tab:ratio_summary}), : %
            \begin{itemize}
                \item $w_{g\gamma}^X/w_{g\Gamma}^X$, or the ratio of the density-shape correlation of shape subsample $X$
                to the %
                density-tidal field correlation, where the tidal field is evaluated at the position of shape subsample $X$ (Sec.~\ref{sim-lin-align}). Using linear theory at large scales, we can assume a linear response $\gamma^{\rm I} = A_{\rm I} \Gamma$, and the ratio reduces to simply $A_{\rm I}$. Good fits to a constant across a range of scales support the tidal-alignment paradigm, by which galaxy shapes align linearly with the local tidal field. %
                \item $w_{g\Gamma}^X/w_{g\Gamma}$, or the density-shape correlations of the galaxy shape subsample $X$'s local tidal field over density-shape correlations of the background tidal field (Sec. ~\ref{sim-dwtf}). %
                Using linear theory at large scales, this ratio can be written as $1 + \mean{\widehat\delta_g\delta_s{\Gamma}}/\mean{\widehat\delta_g{\Gamma}}$.
                Good fits of this ratio to a constant %
                express that the density-weighted tidal field correlations are proportional to the background tidal field correlations and that shape-density clustering is a sub-dominant contribution.

                \item $w_{g\gamma}^X/w_{g\Gamma}$, the ratio of density-shape correlations of the shape subsample $X$ to the correlations of the density tracers and the background tidal field (Sec. ~\ref{sec:sim_dw}). Again using linear theory at large scales and assuming a linear response $\gamma^{\rm I} = A_{\rm I} \Gamma$, the ratio reduces to $A_{\rm I}(1 + \mean{\widehat\delta_g\delta_s{\Gamma}}/\mean{\widehat\delta_g{\Gamma}}$). Good fits to a constant ratio indicate that the non-linear tidal field is aligned in a statistically similar way as the density-weighted shape measurements, where the fit's amplitude quantifies the contribution of density weighting to the galaxy's alignment to cosmic tidal fields.  %
                \item Finally, we compute the goodness-of-fit for ratios relating the density-shape correlations to the density-density correlations as $w_{g\gamma}^X/(w_{gg}^X + 2\Pi_{\mathrm{max}})$ to understand the impact of clustering (Sec.~\ref{sim-clustering}). This ratio measures the average alignment per galaxy pair for a given projected separation. We include the factor of $2 \Pi_{\mathrm{max}}$ to relate the number density, not the excess number density, to the shape correlations %
                \citep{Blazek2015}. Changes in the amplitude of this ratio as $r_p$ decreases inform whether galaxy shapes become more highly aligned or simply are in regions of high galaxy number density relative to larger scales. The large-scale limit of this correlation goes as $\frac{A_{\rm I}}{2 \Pi_{\rm max}}\left(\mean{\widehat \delta_g \Gamma} + \mean{\widehat\delta_g\delta_s{\Gamma}}\right)$.

            \end{itemize}

\begin{table*}
    \centering
    \begin{tabular}{l p{5cm} l}
        \hline
         Ratio&  Assesses& Linear Theory Limit\\\hline
        $w_{g\gamma}^X/w_{g\Gamma}^X$ &  Tidal Alignment& $ A_{\rm I}$\\
       $w_{g\Gamma}^X/w_{g\Gamma}$  &  Tidal Field response to Density Weighting& $1 + \frac{\mean{\widehat\delta_g \delta_X \Gamma}}{\mean{\widehat\delta_g \Gamma}}$\\
       $w_{g\gamma}^X/w_{g\Gamma}$  &  Galaxy Shape response to Density Weighting and Linear Alignment & $A_{\rm I}\left(1 + \frac{\mean{\widehat\delta_g \delta_X \Gamma}}{\mean{\widehat\delta_g \Gamma}}\right)$\\
         $w_{g\gamma}^X/(w_{gg}^X + 2\Pi_{\mathrm{max}})$&  Average Per-Galaxy Alignment& $\frac{A_{\rm I}}{2 \Pi_{\rm max}}\left(\mean{\widehat \delta_g \Gamma} + \mean{\widehat\delta_g\delta_s{\Gamma}}\right)$\\\hline
    \end{tabular}
    \caption{Summary of the different ratios of correlation functions, what they probe, and the expected large-scale limit from linear perturbation theory. %
    }
    \label{tab:ratio_summary}
\end{table*}

            We compute goodness-of-fit tests to a constant across a range of projected radii, $r_p$, of $\wgp$ ratios for either all measured $r_p$ or for larger scales where $r_p\geq1 \, \rm{Mpc}/h$. We bin our data into 12 logarithmically-spaced bins from $0.1 \, \rm{Mpc}/h \leq r_p \leq 20 \,\rm{Mpc}/h$, integrating over 20 linearly-spaced bins, $\Pi$, along the line-of-sight with  $\Pi_{\mathrm{max}} = 20 \, \rm{Mpc}/h$. In practice, we perform all fits omitting the correlations measured at the largest bin of $r_p \approx 16 \, \mpch$. We find that we are no longer sensitive to the statistical properties of galaxy correlations at the largest binned value of $r_p$, but rather become sensitive to the asymmetric large-scale distribution of matter of the total simulation box and periodic boundary effects (Appendix \ref{apx:largescales}). %
            We also omit the smallest two bins of $r_p$, as these are heavily influenced by the smoothing of the tidal field. The jackknife covariance computation follows Eqn.~\eqref{eqn:cov}, where each $\psi_i^{m}$ is the ratio of the \wgp ~measurements for a given jackknife region.

            Our ratio tests for the stellar mass-binned and morphology-binned samples are shown in Figure~\ref{F:all_gals}, while the ratio tests for central galaxies with similar selections are shown in Figure~\ref{F:cen_gals}. %
            In addition to fitting for a constant amplitude, $B$, we also consider a broken power-law of the form %
            \begin{align}  
                B(r_p) = \begin{cases}
                            B & \, r_p > 2 \mathrm{Mpc}/h \\ B \times \left ( \frac{r_p}{2 \mathrm{Mpc}/h} \right )^{\alpha} & r_p \leq 2 \mathrm{Mpc}/h
                            \end{cases}
            \label{eq:plaw}
            \end{align}
            which allows additional scale-dependent freedom to the fit in the 1-halo regime. %
            
         In order to statistically compare two models, we employ the $\chi^2$ difference test, defined as the difference of the best fit $\chi^2$ values divided by the difference of the degrees of freedom, $\nu$, of the two models:
         \begin{equation}
             \Delta \chi^2_\nu = \frac{\chi^2_\mathrm{A} -  \chi^2_\mathrm{B}}{\nu_\mathrm{A} - \nu_\mathrm{B}},
         \end{equation}
         where model B has a greater number of model parameters than model A and both models are fit over the same data. $\Delta \chi^2_\nu=$ 0 when the more complicated model (e.g., a broken power-law model) offers no advantage over the model with fewer parameters (e.g., a constant amplitude model). In practice, increased model flexibility often offers better fits and small values of $\Delta \chi^2_\nu$, but can be interpreted as providing no statistical preference for the more complex model. Conversely, large $\Delta \chi^2_\nu$, along with its $p$-value, can be used to interpret how much a more complicated model is preferred over the simpler one \citep[as used to compare NLA and TATT in][]{campos-ia}.
         We calculate the $\chi^2$ and $p$-value of the fits to these ratios, as well as the $\Delta \chi^2_\nu$ between constant-amplitude fits and broken power-law fits using data across all scales. Given the number of ratios we compute and the noisiness of the data, we adopt $p \lesssim 0.01$ as the threshold to denote a problematic or poor fit.

    \subsection{Density-weighted Calculations for SDSS LOWZ data}\label{sdss-estim} 

        Investigating the impact of density weighting is simpler in hydrodynamical simulations, for which we have the true information of the underlying tidal field. We can create an analog of a density-weighted shear field with SDSS cross-correlations, though, by using Healpix \citep{healpix} maps of the lens and shear sub-samples. We divide the BOSS LOWZ sample into narrow lens redshift bins of $\Delta z \sim 0.02$ and our shape sample into larger bins of $\Delta z \sim 0.1$.%
        We calculate the average galaxy shear and number density in each pixel, as
            \begin{eqnarray}
                 \gamma_P(\vec{\theta})&=\frac{\sum w_{g\in P(\vec{\theta})}\gamma_{g\in P(\vec{\theta})}}{\sum w_{g\in P(\vec{\theta})}}\label{eq:gamma_map}\\
                 1+\delta_g (\vec{\theta}) &=\frac{n_{g\in P}(\vec{\theta})}{\mean{n_g}} .
            \end{eqnarray}
            In the above equation, $\mean{n_g}$ is the average number of galaxies in all pixels and $\vec \theta$ denotes the angular position of a pixel on the sky. %
            We smooth the maps of $n_\text{side}=1024$ with a Gaussian smoothing filter in Fourier space, $S(k)\propto\text{exp}[-(k/k_\text{s})^2]$, before converting back to real space. We choose a smoothing scale of $k_\text{s} = 1~h^{-1} \text{Mpc}$, motivated by the low number density of the lens and source samples, to produce reasonable signal-to-noise when cross-correlating maps.

        We multiply the smoothed shear map by the density map of the LOWZ lens galaxies to obtain density-weighted tidal fields, where we multiply by $\Sigma_c$ to ensure the correct units of the tidal field. We then cross-correlate the lens galaxies with the smoothed tidal-field shear map, both with and without density weighting, to assess the impact of density weighting. The source redshift assigned to each map is the effective redshift of the source subsample that was used in generating the initial map.         

        The low densities of the lens and source samples lead to noisy density-weighted measurements. For higher precision data, we will need to more carefully address systematics that will impact our assembly of the estimators. 
        Since the shear samples typically only have photometric redshifts, the cross correlation measurements are biased. For \DS-like measurements, we typically estimate this bias using a spectroscopic training sample and apply corrections to \DS\ \citep[see for example][]{Blazek2012,probing-gravity}. Since our primary goal in this work is to demonstrate the comparison between density-weighted and un-weighted measurements, we will ignore the subtleties in computing the photo-$z$ corrections. Instead we will directly compare the \DS\ measurements done using pixelized maps with and without density weighting, and compare these measurements. This ensures that within narrow lens bins, the bias due to photo-$z$ error in the two measurements is the same. When combining the lens bins, the bias will differ as contributions from different lens redshift bins scale as $p(z_l)D(z_l)^2 b(z_l)$ for unweighted measurement and  $p(z_l)D(z_l)^2 b(z_l) (1+D(z_l)b(z_l))$ for density weighted measurement \citep[see Eqn.~D10 in][ for calculation of the bias in \DS]{probing-gravity}. We expect this difference to be small in calculating the bias in \DS\ and the final effect in weighted and un-weighted \DS\ comparison to be $\lesssim1\%$ (photo-$z$ bias in \DS\ is $\sim10\%$ and 10\% error in this bias translates to $\sim 1\%$ error in \DS).

        More optimal methods exist to generate the shear map \citep[see discussion in e.g.][]{Singh2021b} and source density-weighting can improve the signal-to-noise ratio of our measurements. However, we leave optimization of these estimators to future work.

    \section{Simulation Results}\label{tng-2pt}

            In this section, we discuss the measurements of the $\wgp$ signal in the TNG100-1 simulation. We maintain the same galaxy density tracer population for all $\wgp$ measurements to be galaxies with stellar mass $M_{*} \geq 10^{9}h^{-1}M_{\odot}$, where the consistent density tracer population enables us to better isolate the effect of galaxy properties on the alignment signal. %

\begin{figure*} 
\begin{center}
\includegraphics[width=1\linewidth]{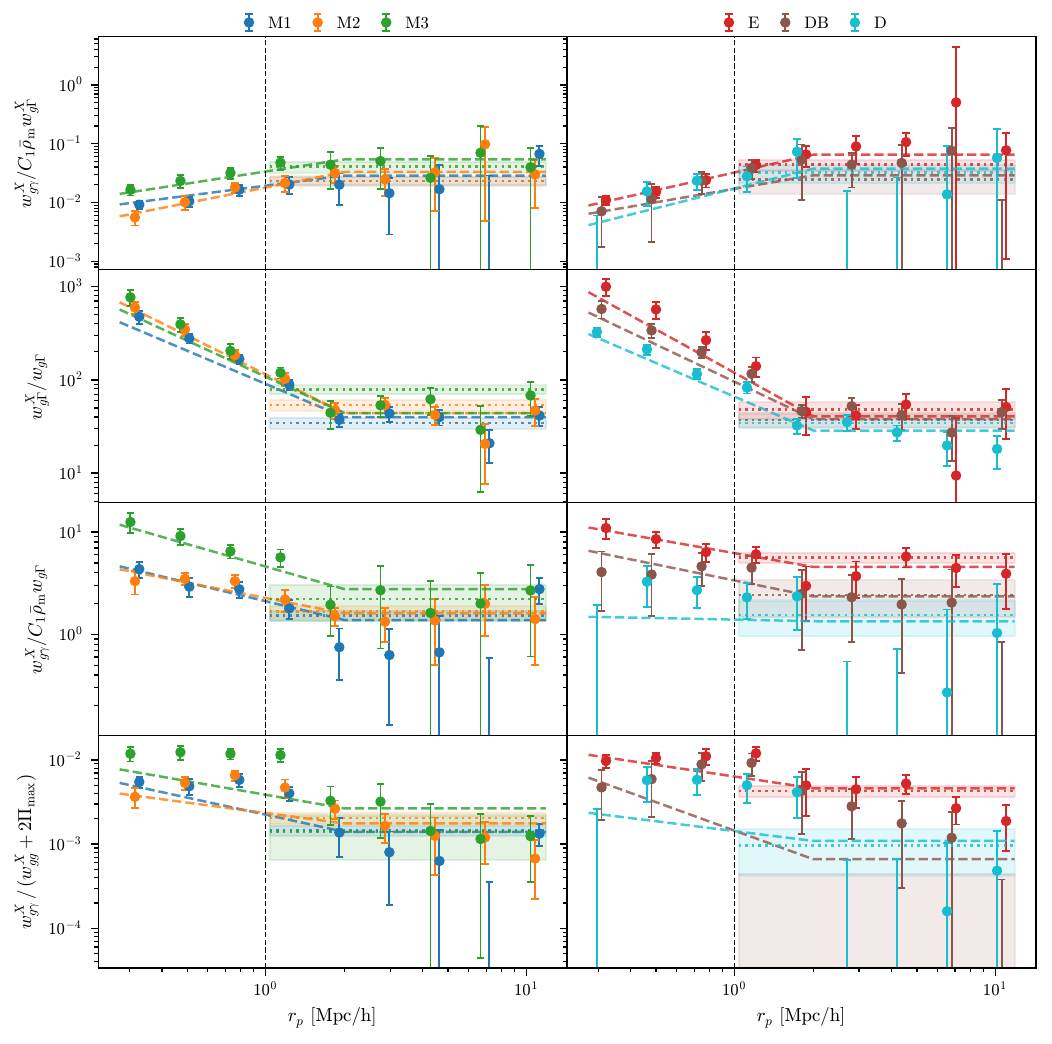}
    \caption{
    Measured projected correlations of the density tracer galaxy sample to the background projected tidal field, $w_g\Gamma$, and galaxy subsamples' shape and tidal field in the IllustrisTNG-100 simulation. Fits to a constant over $r_p \geq 1 \mathrm{Mpc}/h$ are denoted with dotted lines and shaded regions indicating the $1\sigma$ uncertainty of the fit, and broken power-law fits over all scales are denoted with dashed lines. We display only the data used to perform our ratio fits, with the data points slightly offset horizontally for readability. %
    \label{F:all_gals}}
\end{center}
\end{figure*}

\begin{figure*} 
\begin{center}
\includegraphics[width=1\linewidth]{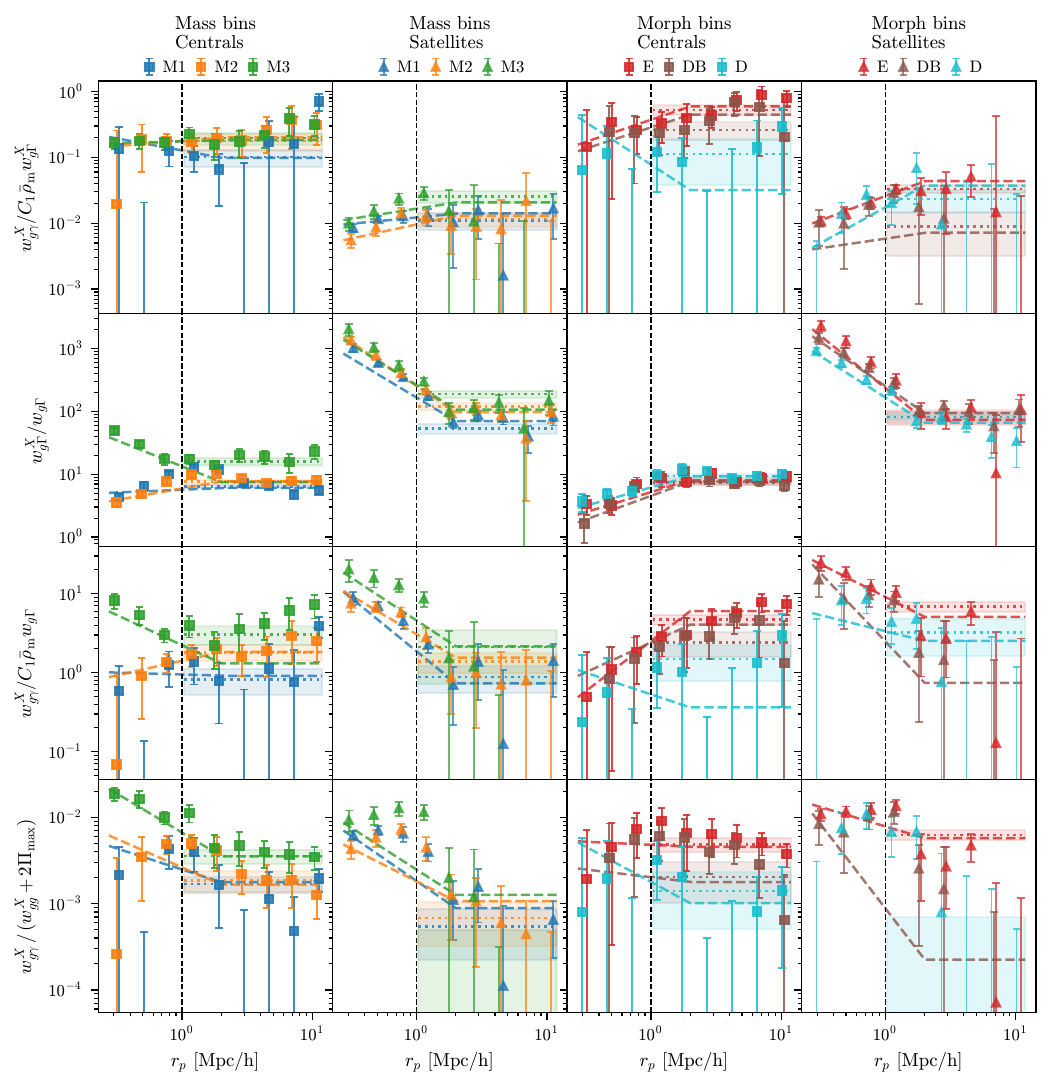}
\caption{\label{F:cen_gals} Same as Figure~\ref{F:all_gals}, but for splits between central and satellite galaxy samples. Unlike in Figure \ref{F:all_gals}, we find acceptable fits to a constant ratio in the first row of panels across the considered $r_p$ for all central mass and morphology bins, indicating that central galaxies are tidally aligned even at small scales. Additionally, we find good constant fits in the 1-halo regime when considering density-weighted alignments in the third row of panels. We also find statistically significant large-scale amplitude hierarchies between the lowest and highest mass bins, and between elliptical and disc galaxies. Satellite galaxies have much larger tidal field amplitudes and live in more crowded environments, leading to large differences in the amplitudes of the ratios relative to the central samples.   %
}
\end{center}
\end{figure*}

            \subsection{Evaluating the Tidal-Alignment Paradigm} \label{sim-lin-align}

           We begin by examining galaxy shapes and their alignment with the underlying tidal field at the galaxy's position by examining $w_{g\gamma}^X/w_{g\Gamma}^X$ ratios. %
           Some galaxies are postulated to align linearly with the matter tidal field at a given position \citep{Catelan2001}, a statement that we can directly probe through hydrodynamical simulations. Good fits of the ratio to a constant indicate that the galaxy shapes are tidally aligned to their underlying field within our error bars. %

We find that a constant is a good fit to all galaxy samples over scales $r_p \geq 1 \mathrm{Mpc}/h$ %
and remains valid for the M1, E, DB, and D samples across all $r_p$ considered in our analysis (first row of Figure~\ref{F:all_gals}, Table~\ref{tab:all_const_fits}). A feature in the ratio appears in the 1-halo regime, where the ratio's amplitude decreases, reducing the goodness of fit for some samples. Allowing additional freedom in the fit through a broken power law (Eqn.~\ref{eq:plaw}) is preferred across all scales relative to a constant fit ($\Delta \chi^2 = 2.4 - 20.0$) and is statistically consistent with the amplitude found from the constant fits over $r_p \geq 1 \mathrm{Mpc}/h$. This indicates that some feature of galaxy shapes or tidal fields in the 1-halo regime does not generally align with the expectations of the linear alignment paradigm.

	We further subdivide our samples into central and satellite populations to better understand the physics in the 1-halo regime that breaks linear tidal alignment. For the central galaxy sub-samples (First row of Fig.~\ref{F:cen_gals}, Table~\ref{tab:cen_const_fits}), we find good fits to a constant amplitude across all scales (except for the M1 sub-sample). There is now no model preference for adding scale-dependent freedom via Eqn.~\eqref{eq:plaw}. The satellite sub-samples (not shown) only yield good all-scale fits to a constant for the D, DB, and M1 samples, where additional freedom via Eqn.~\eqref{eq:plaw} does not generally improve the goodness-of-fit. 
    
	We plot histograms of the central and satellite smoothed projected tidal field ellipticities against the smoothed local matter density fluctuation in Figures~\ref{fig:d_e_stellar} and~\ref{fig:d_e_morph} to better understand the alignment between the tidal field at galaxy positions and the galaxy shapes. %
    We find that central galaxies live in tidal fields with much lower tidal field ellipticity than satellite galaxies. Satellites live in their own subhalos and where their alignments respond to higher-order tidal effects. The projected tidal field ellipticity can be much larger (as discussed in Section \ref{tidal-shear},  %
    Eqn.~\ref{eqn:smoothed-tidal-field}), resulting in much lower alignment amplitudes for satellites than for central galaxies with like properties over $r_p \geq 1 \mathrm{Mpc}/h$.

From these tests, we find that galaxy shapes align linearly with their underlying tidal field for all sub-samples over $r_p \geq 1 \mathrm{Mpc}/h$ and continue to align well over all scales used in the simulation for central galaxies. Additional effects, such as higher-order alignment, impact this statement in the 1-halo regime when assessing satellite alignment, affecting different mass/morphology samples in ways that depend on their central/satellite mix.

            \subsection{Behavior of the Density-Weighted Tidal Field}\label{sim-dwtf}
Understanding the tidal field alignment at galaxy positions relative to the background tidal field is important for assessing the impact of density weighting on galaxy intrinsic alignments. To characterize the density-weighted tidal field, we perform similar fits as before, now to $w_{g\Gamma}^X/w_{g\Gamma}$ ratios. 

	We find that a constant ratio is insufficient to characterize our samples over both all scales of the analysis and $r_p \geq 1 \mathrm{Mpc}/h$ (Row 2, Figure~\ref{F:all_gals}). Additional scale-dependent freedom through Eqn.~\eqref{eq:plaw} improves all fits across all scales except for the M1 and M3 samples. Interestingly, we find that $\alpha$ is greater than zero by 3 sigma for all samples with acceptable $p$-value fits (M2, E, DB, and D), suggesting a density-weighted dependency of the tidal field in the 1-halo regime (like clustering). We once again separate our samples into central and satellite sub-populations to further probe the 1-halo dependencies.

\begin{figure}
    \centering
    \includegraphics[width=1\linewidth]{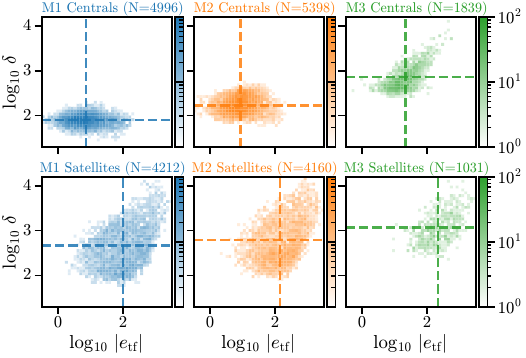}
    \caption{The tidal field ellipticity versus the total matter field density contrast at the positions of galaxies for the stellar mass-binned samples, divided into central and satellite populations. Both the density contrast and the tidal field are smoothed by the Gaussian smoothing kernel of Eqn.~\eqref{eqn:smoothed-tidal-field} for a smoothing scale of 0.1 Mpc/$h$. Dashed lines denote the average log density contrast and log tidal field ellipticity of each sample. For a given mass range, central galaxies are generally located in regions with a lower density contrast and tidal field ellipticity, whereas satellites live in a large range of higher density contrast and tidal field ellipticity.}
    \label{fig:d_e_stellar}
\end{figure}

\begin{figure}
    \centering
    \includegraphics[width=1\linewidth]{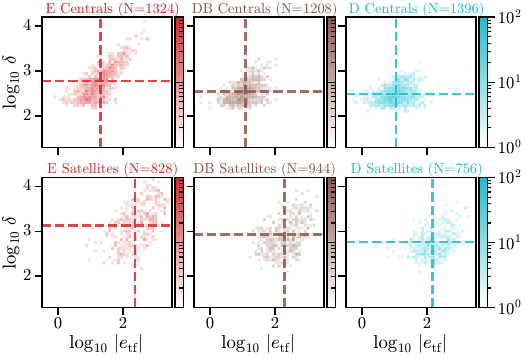}
    \caption{Same as Figure~\ref{fig:d_e_stellar} for the morphology-binned samples. For these sub-samples, satellite galaxies span a large range of higher density contrast and tidal field ellipticity relative to the central sub-samples. Central elliptical galaxies also span a large range, in contrast to the central disc galaxy distribution, which is more concentrated around lower density contrast and tidal field ellipticity.} %
    \label{fig:d_e_morph}
\end{figure}

	For the central galaxy sub-samples, we find that a constant fit is only applicable for the E, D, M2, and M3 samples when $r_p \geq 1 \mathrm{Mpc}/h$ (second row of Fig.~\ref{F:cen_gals}). Only the E satellite sub-sample can be fit to a constant for $r_p \geq 1 \mathrm{Mpc}/h$. Using Eqn.~\eqref{eq:plaw}, we find distinct differences between central and satellite tidal fields when fitting across all scales. The central sub-samples have good fits to the broken power-law for the M2, E, and D samples with $\alpha$>0 by 3 sigma for these fits. On the other hand, the satellite sub-samples (not shown) fit the broken power-law well for all samples except M1, with $\alpha$ < 0 by 3 sigma. Through our fits, we find the relative amplitude of the density-weighted tidal field decreases with respect to the background at central galaxy locations as $r_p$ decreases, but increases at satellite galaxy locations. This feature is likely due to the projected tidal field at satellite locations being of larger amplitude and less isotropic than at central galaxy locations (shown in Figures~\ref{fig:d_e_stellar} and~\ref{fig:d_e_morph}).

            \subsection{Density-Weighted Galaxy Shape Alignments} \label{sec:sim_dw}
            We now assess how the density-weighted shape alignment of galaxies relates to the background tidal field by examining $w_{g\gamma}^X/w_{g\Gamma}$ ratios. Good constant-amplitude fits indicate that galaxy shapes are linearly aligned with the background tidal field (Row 3, Figure~\ref{F:all_gals}). We find fine constant-amplitude fits for the E, DB, D, and M2 samples when $r_p \geq 1 \mathrm{Mpc}/h$ and for DB, D, and M2 over all scales. At large scales, the density-weighting contribution increases the amplitude of the fit relative to $w_{g\gamma}^X/w_{g\Gamma}^X$ ratios, in line with expectations that the linear bias of the tracer sample modulates the large-scale alignment amplitude (see Table~\ref{tab:all_const_fits}). Deviations from a constant at small scales indicate higher-order alignment and/or galaxy clustering are influencing the galaxy alignment strength.
            In fact, we find power-law fits over all scales improve these ratio fits for all samples except M1, where the M2, M3, and E samples indicate an increase in the alignment strength towards small $r_p$ with $\alpha$<0 by 3 sigma. 
            
	Constant-amplitude fits improve when considering centrals over all scales (third row of Fig.~\ref{F:cen_gals}, Table~\ref{tab:cen_const_fits}). All sub-samples yield good fits, and a hierarchy between mass-binned and morphology-binned samples emerges: M3 galaxies have a statistically higher alignment amplitude ($B = 2.53 \pm 0.46$) relative to the M1 galaxies ($B = 0.94 \pm 0.27$), and E galaxies ($B = 3.08 \pm 0.50$) have a much higher amplitude than D galaxies ($B = 0.62 \pm 0.47$, consistent with a very small or null alignment). Adding additional freedom via the power-law fit does not substantially improve the goodness-of-fit or indicate substantial scale dependence in the ratio for centrals (with the exception of the M3 and E samples). The D sub-sample is the only satellite population ratio (not shown) that can be fit to a constant amplitude, where the alignment amplitude is again near-zero ($B =  2.66 \pm 1.43$).

            \subsection{Impact of Non-Linear Galaxy Clustering on Shape Alignments}\label{sim-clustering}
            An important distinction to make now is the source of the increase in alignment for some of the samples discussed in the previous subsection: whether the density weighting contributions come from galaxy clustering or other higher-order tidal field effects. We probe the impact of galaxy clustering on shape alignments by performing fits on $w_{g\gamma}^X/(w_{gg}^X + 2 \Pi_{\mathrm{max}})$ ratios, where we make the distinction to take ratios over the galaxy number density instead of the excess number density. %
            Increases in this ratio as a function of scale indicate that the alignment amplitude is not due to clustering, but due to increased shape alignment from non-linear tidal effects.
            For every sample we examine, we find an increase in this ratio in the 1-halo regime (Row 4, Figure \ref{F:all_gals}), with the relative increase for satellite galaxies much larger than for centrals as separation decreases. %
            
Subdividing into central galaxies, a constant fit is sufficient across all scales for the M1, M2, E, and D samples (fourth row of Fig.~\ref{F:cen_gals}), but the M3 sample shows a definitive increase in the small-scale ratio. This trend suggests that density weighting enhances the 1-halo average per-galaxy alignment for the highest-mass central galaxy sample. We see a similar trend for satellite galaxies, although the increase in average galaxy alignment in the 1-halo regime seems to be more influenced by clustering, where these ratios can be 2-10 times smaller for satellites of a given selection than centrals of the same subsample. %


\section{Measurements in BOSS-LOWZ} \label{boss-2pt}
        	In this section we present the measurements using the spectroscopic galaxies from the BOSS LOWZ sample as described in Sec.~\ref{sec:data_sdss}.

               \begin{figure}
                  \centering
                    \includegraphics[width=\columnwidth]{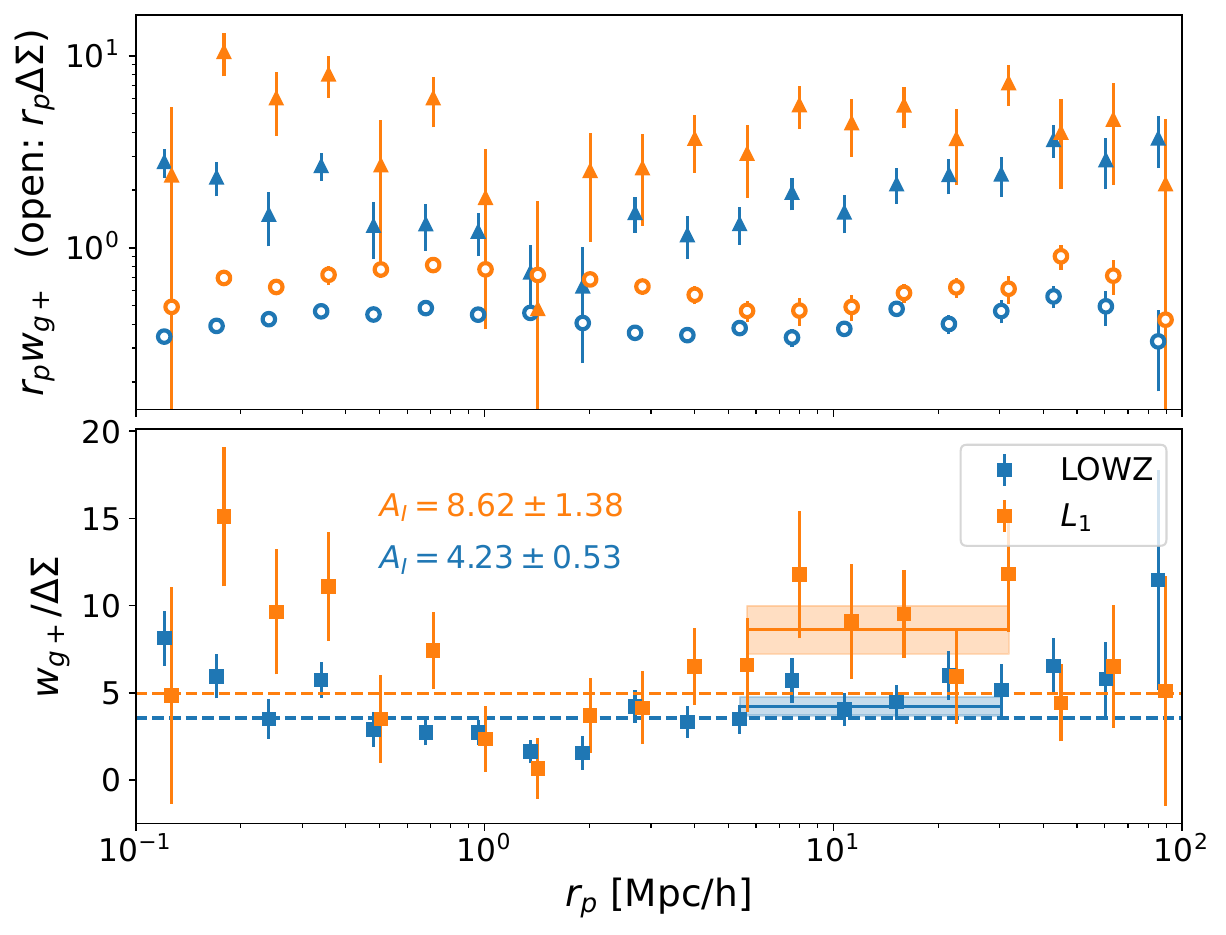}
                    \caption{Comparison of the \wgp\ measurements and the rescaled \DS\ measurements from galaxy-lensing cross correlations. Blue points show measurements 
                    for the full BOSS LOWZ sample, while orange points show the brightest (20 per~cent) subsample. The top panel shows the measurements, while the bottom panel is the 
                    ratio, \wgp/\DS.  %
                    In the lower panel, solid lines show the mean and standard deviation of the ratio in the range $5<r_p<40~\mpch$, while dashed lines show the best fit over all scales. 
                    We find results consistent with the analytical NLA fits from  
                    \protect \cite{Singh2015}, while on small scales there are scale-dependent deviations from the model, which are indicative of the deviations from the NLA model.%
                     }
                     \label{fig:wgp_DS_comp}
              \end{figure}

In Fig.~\ref{fig:wgp_DS_comp} we compare the \wgp\ measurements and the rescaled \DS\ measurements (as explained in Sec.~\ref{sdss-estim}), where good fits to a constant ratio can be interpreted as good fits to the NLA model with a scale-independent 
              amplitude $A_{\rm I}$. 
              On large scales, $r_p\gtrsim 2~\mpch$, the \wgp\ and \DS\ measurements have similar scale dependence and thus \wgp\ can be fit with a simple multiplicative amplitude shift of \DS.
              This finding is consistent with the simulations as well as  the model predictions, where the density weighting effect primarily changes the amplitude of \wgp\ with 
              relatively small deviations in the slope.%
              Fitting over the range $5<r_p<40 h^{-1} \text{Mpc}$, we find $A_{\rm I} = 4.23 \pm 0.53$ for the LOWZ sample and $A_{\rm I} = 8.62 \pm 1.38$ for the 20\% most luminous galaxies in the sample. These fits have $p$-values of 0.55 and 0.66, respectively, and are consistent within $1\sigma$ of previous model-dependent fits on $A_{\rm I}$ \citep{Singh2015}. Extending the fit region to smaller separation scales ($0.1<r_p<40 h^{-1} \text{Mpc}$) do not yield good fits, but extending to larger scales ($5<r_p<100 h^{-1} \text{Mpc}$) are consistent with our previous findings: $A_{\rm I} = 4.26 \pm 0.53$ for the LOWZ sample and $A_{\rm I} = 7.75 \pm 1.32$ for the high-luminosity sample with $p$-values of 0.72 and 0.56, respectively.
 On smaller scales, $r_p\lesssim2\mpch$, we see significant deviations in 
              the slopes of the two signals from their slopes on large scales, suggesting the stronger influence of density weighting effects or other nonlinear effects, like tidal torquing and nonlinear bias. %
              
\begin{figure}
                  \centering
                    \includegraphics[width=\columnwidth]{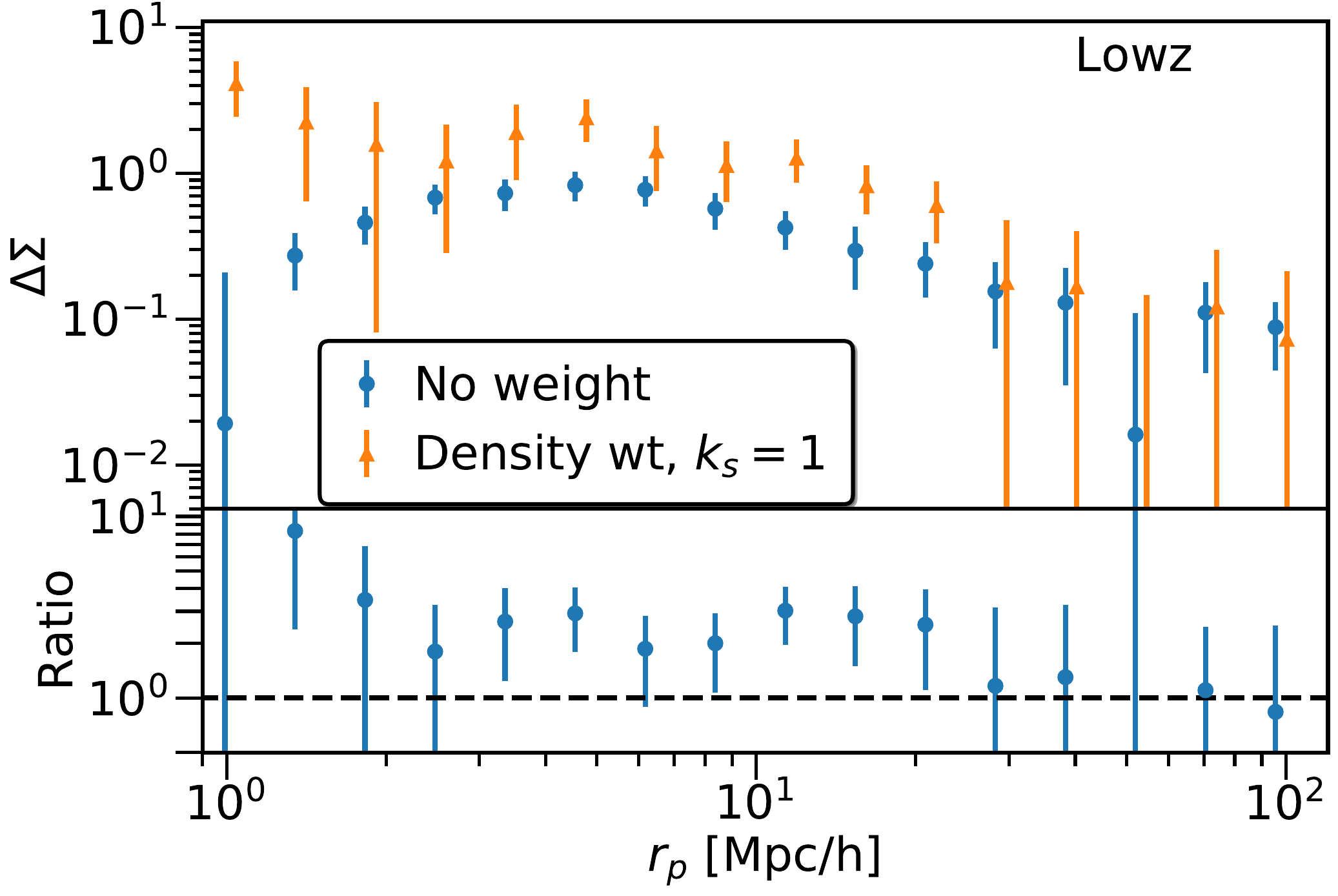}
                    \caption{\label{fig:DS_den_wt} \DS\ measurements with and without the density weighting effects. These measurements are done using SDSS and BOSS LOWZ samples, as in Fig.~\ref{fig:wgp_DS_comp}. For the purpose of marginalizing over the IA effects, the NLA model is likely to suffice over the scales of $\sim$ 2 -- 5 $h^{-1} \text{Mpc}$ given our statistical precision. 
                    }
                     
              \end{figure}

              We reiterate that this rescaled \DS\ in Fig. \ref{fig:wgp_DS_comp} does not include the effects of density weighting on the galaxy shape field, which could lead to differences between the measured \wgp\ on smaller scales where clustering and associated nonlinear bias are stronger.
              The density-weighted analog of \DS\ using our map-based estimates, as described in Sec. \ref{sdss-estim},  is designed to isolate the impact of the tidal field response from these other effects. These results are shown in Fig.~\ref{fig:DS_den_wt}. 
              This comparison suggests that for the purpose of marginalizing over the IA
              effects in the weak lensing measurements, the NLA model is likely to suffice over the scales of $\sim 2-5\mpch$ within our statistical precision. %
              Though, it is still unclear whether NLA will suffice when using information from small scales ($r_p \lesssim$ 2). %

It is important to note that the comparison in Fig.~\ref{fig:wgp_DS_comp} has small biases due to the following differences in the lensing and $\wgp$ estimators: i) different effective redshifts of the measurements, ii) the impact of redshift-space distortions on \wgp\, and iii) differences in the line of sight window for lensing and \wgp\ estimators. These corrections are typically of order 1--2 per cent, similar to the corrections in the $E_G$ estimator that compares lensing and galaxy clustering measurements \citep[see fig.~5 in][]{probing-gravity}. Since our IA amplitude constraints have statistical uncertainties of order 10 per cent, we will ignore these biases  in this work. Future studies with higher precision measurements should study and, if needed, correct for these biases.

              Here we demonstrated a novel methodology to test assumptions of a linear response of galaxy shapes to the tidal fields. 
              As a final note, the demonstrated method could be used on current and forthcoming Stage IV surveys \citep[e.g.][]{2026AJ....171..285D} which should improve statistical precision, in turn giving us deeper insights into the behavior of higher-order terms in tidal alignment models.

\section{Conclusions}\label{conc}
The intrinsic alignments of galaxy shapes are a leading astrophysical systematic in weak gravitational lensing analyses. The two analytic nested models that are commonly used for marginalizing over intrinsic alignments (TATT with 3+ parameters and NLA with 1+ parameter) carry the assumption that intrinsic galaxy shapes respond to tidal fields, with a purely linear response in NLA and additional density weighting and 2nd-order tidal torquing terms in TATT. While many analyses have fit these (and other) models to two-point functions, these studies are limited by the precision of the measurements and the analytic models' limited ability to describe the physical mechanisms that source the galaxy intrinsic alignments.  %

In this work, we explored model-independent methods of assessing the role of tidal alignments in galaxy intrinsic alignments %
 and the dependencies of density-weighting through proxy models of two-point correlations, both in simulations and real data, that implicitly incorporate non-linear clustering.
 In simulations, we assess the impact of density-weighting and the fidelity of the tidal alignment model by relating ratios of two-point functions over a variety of galaxy properties by leveraging the direct access to tidal-field information.
We find the tidal alignment paradigm sufficiently characterizes galaxy alignments in IllustrisTNG at scales larger than 1 \mpch, both with and without density-weighting contributions. The alignment amplitude of central galaxies increases with stellar mass and as galaxies become more bulge-dominated; in fact, within the precision of our measurements, most density-shape correlations of central galaxies match the trends of the associated density-tidal field correlations at separation scales well into the 1-halo regime.
On the other hand, the density-shape correlation of disc galaxies is weakly associated with the density-shape correlations of the underlying tidal field, where we find fits to a constant amplitude relating them to be consistent with zero.
The average per-galaxy shape alignment of our subsamples generally increased towards smaller scales, with the largest mass galaxies increasing more than the low-mass subsamples. For centrals, the largest-mass sample shows a clear increase in average per-galaxy alignment at small separation scales, with a less pronounced increase in the lower-mass bins. %
This trend indicates that increases in intrinsic alignment two-point functions on small scales beyond those expected for NLA are not driven exclusively by nonlinear clustering. 
The same metric for the satellite subsamples shows a large relative increase in the 1-halo regime, but with a lower amplitude than the central galaxy subsamples.

We examined the underlying amplitude of the smoothed tidal field against the smoothed density contrast at galaxy locations. %
We generally find that satellite galaxies span a broader range in matter density contrast and reside in environments with larger tidal field amplitudes than centrals with the same mass or morphology selection. Central galaxies live in potential minima at the center of a host halo, but satellites live in subhalos and respond to the tidal field in a scale-dependent way due to density-weighting contributions, creating a range of possibilities in these evolving environments. %
These trends support our conclusions on central galaxy alignments, showing that the underlying fields at central galaxy locations can be characterized more cleanly/uniformly than at satellite locations.
We note that the precision of our measurements suffers from limited statistical power of the simulation volume and the number of galaxies in our sample. Future examinations of the linear alignment paradigm using higher-precision simulations in larger box sizes can assess the applicability of our findings to future surveys.

Our model-independent investigations of the IllustrisTNG hydrodynamical simulation reinforce the findings found in the BOSS LOWZ data: Galaxies align linearly with respect to the underlying tidal field, and the density-weighting contribution manifests as an increase in the alignment strength (at sufficiently large $r_p$). %
We find that for both simulations and real data, the two proxy models (from the tidal shear field) show similar scale dependence of alignment on large scales ($\geq 2h^{-1}$Mpc), fitting well to a constant amplitude. The density weighting term mainly changes the amplitude of the signal, which is due to the higher weight placed on regions of large density and thus environments with large tidal fields. %
Applying our model-independent investigation can lead to more informative results using the latest spectroscopic measurements from DESI, where the increased density of observations will improve density-weighted shear maps. 

In future work, we would like to further investigate the torquing mechanism of the tidal fields and its interplay with the disc-dominated galaxies' internal angular momentum, as it can provide insight into why disc-dominated galaxies do not exhibit a measurable IA signal on large scales.%
While definitive trends are limited by the statistics of our samples, an interesting extension would be to further test these relationships with 3D two-point functions and at the field level (e.g., \citealt{samuroff-2020}) or with estimators that increase the signal of galaxy alignments.  This could include, for example, multipole estimators \citep{shakir}, multiplet estimators \citep{2024MNRAS.534.3540L}, large-scale axial intrinsic alignment estimators \citep{2025arXiv251110005D}, and other estimators that are more optimally-weighted than the projected statistics used here \citep{Lamman2025Optimal}. %
Nevertheless, the work we present enables data-informed model and sample selection by examining the underlying physical mechanisms for IA, characterizing one of the largest sources of systematic uncertainty for upcoming Stage-IV cosmological inferences of weak lensing. %
Applications of our methodology to larger-volume simulations and to more-constraining datasets will place even stronger constraints on both models and the underlying processes of galaxy intrinsic alignments. %

\section*{Data Availability Statement}

The simulation catalog data is available at \url{https://github.com/McWilliamsCenter/gal_decomp_paper}, while the SDSS-III BOSS LOWZ catalog is available at \url{https://cosmo.vera.psc.edu/SDSS_shape_catalog/}.  Random catalogs for the LOWZ sample can be downloaded from the SDSS-III data release site, \url{https://data.sdss.org/sas/dr12/boss/lss/} (DR12 v5).
The software used to calculate the correlation functions in this work is available at \url{https://github.com/sukhdeep2/corr_pc}. %

\section*{Acknowledgments}
The authors thank Yesukhei Jagvaral for sharing his catalogs and for his initiative in accelerating the early work of this project.
PR and RM were supported in part by Department of Energy grant
DE-SC0010118 and in part by a grant from the Simons Foundation (Simons
Investigator in Astrophysics, Award ID 620789). SS was supported by the McWilliams postdoctoral fellowship at CMU.
JB was supported in part by the National Science Foundation (AST-2206563, AST-2442796), the Department of Energy (DE-SC0024787), and NASA (Roman Research and Support Participation program, 80NSSC24K0088). This work benefited from the resources and support provided by the echo-IA collaboration (https://echo-ia.org).

\bibliographystyle{mnras}
\bibliography{BIB, IA} %

\appendix

\section{Effects of smoothing scales of the tidal field on the measurements.}\label{apx:smooth}
\begin{figure}
     \centering
     \includegraphics[width=1\linewidth]{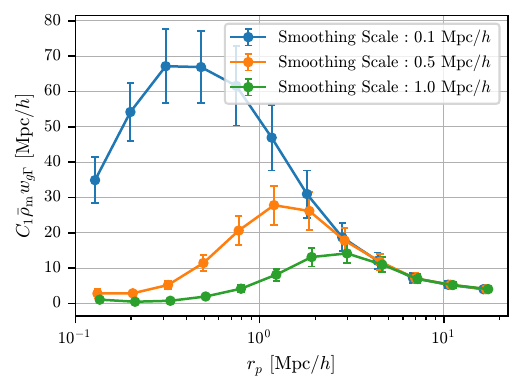}
     \caption{The $w_{g\Gamma}$ correlations between galaxy positions and the tidal fields measured at random locations for different smoothing scales of the tidal field. All correlations converge at separation scales much larger than the smoothing scale and diverge at approximately three times the smoothing scale.}
     \label{fig:smooth_wgp}
 \end{figure}

In this short appendix, we show the results of smoothing the projected tidal field on our measurements of $w_{g\Gamma}$ in TNG100-1. We consider three different smoothing scales by setting $r_\mathrm{s}$ in Eqn.~\eqref{eqn:smoothed-tidal-field} to 0.1, 0.5, or 1.0 Mpc/$h$. The results are plotted in Figure \ref{fig:smooth_wgp}, where we see qualitative agreement with previous studies \citep[e.g.][]{Chisari_2013}. At approximately three times the smoothing scale, we observe a turnover in $w_{g\Gamma}$ that indicates the smoothing of the tidal field marginalizes over the structure within this range. This smoothing feature determines which small scales we exclude from our analysis. At large scales, all correlations coincide with one another. The smoothing scale can therefore be set to relate galaxy shapes to the orientation of structures at scales a few times larger than the smoothing scale \citep[as discussed in][]{Blazek2015, zjupa-ia}.

\section{Large-Scale Scale-Cuts in Simulation}
\label{apx:largescales}
In this appendix, we discuss excluding correlation functions at the largest $r_p$ in our simulation volume. As stated in Section \ref{tng-2pt}, we evaluate $w_{g+}$ correlation functions at 12 log-spaced bin of $r_p$ spanning from 0.1-20 Mpc/$h$. While the two bins at the smallest $r_p$ are omitted due to the smoothing of the tidal field, we also omit the largest $r_p$ bin due to asymmetries when integrating along the line-of-sight. These large-scale asymmetries were first noted in the TNG100-1 box in \cite{zjupa-ia} via galaxy mass selections in 1-point statistics, where we now quantify this property for two-point statistics.

\begin{figure}
    \centering
    \includegraphics[width=1\linewidth]{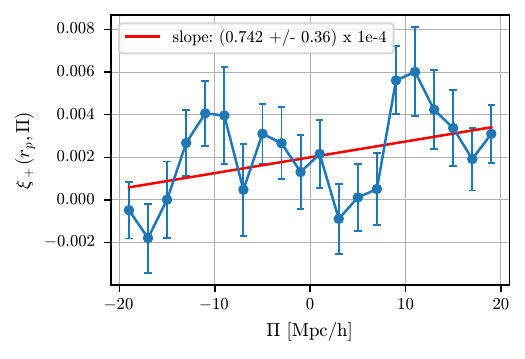}
    \caption{Eqn.~\eqref{eq:xigp} of the density-shape correlations for the M3 sample at $r_p \approx 16 ~\mathrm{Mpc}/h$. Without volume-limiting effects or cosmic variance, the correlations about $\Pi=0~\mathrm{Mpc}/h$ should be symmetric, corresponding to a fitted slope consistent with zero. For this largest bin, we find the slope to be over $2\sigma$ discrepant from zero. We omit the largest bin in our analysis to remove this feature from our correlations and ratio fits. %
    }
    \label{fig:asym}
\end{figure}

At the largest %
scales we consider, we find we are susceptible to cosmic variance in the realization of the box and found by our integral along the line-of-sight in $\Pi$ (Eqn.~\ref{eqn:wgp}), where the statistical distribution of galaxies about $\Pi=0 ~\mathrm{Mpc}/h$ is not symmetric. We quantify the degree of asymmetry by fitting a line to the data and measure the deviation of the slope from zero using the uncertainty of the fit, where we omit data points that deviate from a zero slope by over $2\sigma$. We find for some samples (e.g., the M3 sample) that our statistics are skewed as we approach the volume limit of the simulation box (shown in Figure \ref{fig:asym}), where the slope of the fitted line exceeds our threshold. We omit these scales from the analysis for all sub-samples in order to make more direct comparisons between the statistics of each sample.

\section{Tables of Fits}
\label{apx:tables}
Best-fit values and uncertainties are provided here for each sample and sub-sample of the TNG100-1 ratios, specifically for the $w_{g\gamma}^X/w_{g\Gamma}^X$ and $w_{g\gamma}^X/w_{g\Gamma}$ ratios, along with their goodness-of-fit criteria. %

\begin{table*}
\centering
\caption{Constant first across large scales and all scales considered in TNG100-1 for different shape-to-tidal-field ratios of \wgp. The relevant prefactors of $C_1 \bar{\rho}_{\rm m}$ are included in the fits of galaxy-tidal field correlations but are omitted from the tables for compactness.}
\label{tab:all_const_fits}
\begin{tabular}{lcc}
\hline\hline
Ratio & Large-scale: $B \pm \sigma_B\ (\chi^2(\nu)),\ p$ & All-scale: $B \pm \sigma_B\ (\chi^2(\nu)),\ p$ \\
\hline
$w_{g\gamma}^{M1}/w_{g\Gamma}^{M1}$ & $0.0233 \pm 0.0035\ (6.37\,(5)),\ 0.2716$ & $0.0115 \pm 0.0012\ (16.10\,(8)),\ 0.0410$ \\
$w_{g\gamma}^{M2}/w_{g\Gamma}^{M2}$ & $0.0232 \pm 0.0045\ (1.04\,(5)),\ 0.9589$ & $0.0080 \pm 0.0011\ (28.51\,(8)),\ 0.0004$ \\
$w_{g\gamma}^{M3}/w_{g\Gamma}^{M3}$ & $0.0408 \pm 0.0086\ (2.55\,(5)),\ 0.7690$ & $0.0135 \pm 0.0025\ (23.14\,(8)),\ 0.0032$ \\
$w_{g\gamma}^{M1}/w_{g\Gamma}$ & $1.5303 \pm 0.1725\ (24.46\,(5)),\ 0.0002$ & $1.7728 \pm 0.1584\ (35.22\,(8)),\ 0.0000$ \\
$w_{g\gamma}^{M2}/w_{g\Gamma}$ & $1.6724 \pm 0.2146\ (2.59\,(5)),\ 0.7623$ & $2.0719 \pm 0.1873\ (19.29\,(8)),\ 0.0134$ \\
$w_{g\gamma}^{M3}/w_{g\Gamma}$ & $2.2182 \pm 0.8528\ (28.09\,(5)),\ 0.0000$ & $3.6605 \pm 0.6693\ (32.77\,(8)),\ 0.0001$ \\
$w_{g\gamma}^{E}/w_{g\Gamma}^{E}$ & $0.0439 \pm 0.0091\ (3.15\,(5)),\ 0.6771$ & $0.0086 \pm 0.0013\ (19.50\,(8)),\ 0.0124$ \\
$w_{g\gamma}^{DB}/w_{g\Gamma}^{DB}$ & $0.0247 \pm 0.0105\ (5.28\,(5)),\ 0.3821$ & $0.0062 \pm 0.0038\ (8.36\,(8)),\ 0.3993$ \\
$w_{g\gamma}^{D}/w_{g\Gamma}^{D}$ & $0.0325 \pm 0.0114\ (4.76\,(5)),\ 0.4453$ & $0.0079 \pm 0.0035\ (10.90\,(8)),\ 0.2077$ \\
$w_{g\gamma}^{E}/w_{g\Gamma}$ & $5.6330 \pm 0.6578\ (11.02\,(5)),\ 0.0510$ & $6.2092 \pm 0.5623\ (22.50\,(8)),\ 0.0041$ \\
$w_{g\gamma}^{DB}/w_{g\Gamma}$ & $2.4216 \pm 0.9505\ (9.15\,(5)),\ 0.1032$ & $3.2084 \pm 0.8177\ (11.08\,(8)),\ 0.1974$ \\
$w_{g\gamma}^{D}/w_{g\Gamma}$ & $1.5492 \pm 0.5930\ (7.65\,(5)),\ 0.1768$ & $1.3788 \pm 0.4390\ (8.95\,(8)),\ 0.3469$ \\
\hline
\end{tabular}
\end{table*}

\begin{table*}
\centering
\caption{Same as Table \ref{tab:all_const_fits}, but only for central galaxies.}
\label{tab:cen_const_fits}
\begin{tabular}{lcc}
\hline\hline
Ratio & Large-scale: $B \pm \sigma_B\ (\chi^2(\nu)),\ p$ & All-scale: $B \pm \sigma_B\ (\chi^2(\nu)),\ p$ \\
\hline
$w_{g\gamma}^{M1\,({\rm cen})}/w_{g\Gamma}^{M1\,({\rm cen})}$ & $0.1006 \pm 0.0295\ (16.57\,(5)),\ 0.0054$ & $0.1181 \pm 0.0275\ (20.51\,(8)),\ 0.0086$ \\
$w_{g\gamma}^{M2\,({\rm cen})}/w_{g\Gamma}^{M2\,({\rm cen})}$ & $0.1769 \pm 0.0510\ (1.34\,(5)),\ 0.9303$ & $0.1790 \pm 0.0406\ (2.22\,(8)),\ 0.9734$ \\
$w_{g\gamma}^{M3\,({\rm cen})}/w_{g\Gamma}^{M3\,({\rm cen})}$ & $0.1972 \pm 0.0409\ (3.36\,(5)),\ 0.6441$ & $0.1683 \pm 0.0210\ (4.23\,(8)),\ 0.8357$ \\
$w_{g\gamma}^{M1\,({\rm cen})}/w_{g\Gamma}$ & $0.8230 \pm 0.3010\ (14.40\,(5)),\ 0.0133$ & $0.9365 \pm 0.2733\ (16.84\,(8)),\ 0.0318$ \\
$w_{g\gamma}^{M2\,({\rm cen})}/w_{g\Gamma}$ & $1.8035 \pm 0.4082\ (1.07\,(5)),\ 0.9563$ & $1.3609 \pm 0.2772\ (4.14\,(8)),\ 0.8443$ \\
$w_{g\gamma}^{M3\,({\rm cen})}/w_{g\Gamma}$ & $3.0409 \pm 0.8047\ (4.74\,(5)),\ 0.4486$ & $2.5384 \pm 0.4583\ (12.94\,(8)),\ 0.1141$ \\
$w_{g\gamma}^{E\,({\rm cen})}/w_{g\Gamma}^{E\,({\rm cen})}$ & $0.5261 \pm 0.0921\ (4.08\,(5)),\ 0.5377$ & $0.4600 \pm 0.0815\ (6.34\,(8)),\ 0.6090$ \\
$w_{g\gamma}^{DB\,({\rm cen})}/w_{g\Gamma}^{DB\,({\rm cen})}$ & $0.2631 \pm 0.0775\ (3.52\,(5)),\ 0.6208$ & $0.3214 \pm 0.0696\ (8.87\,(8)),\ 0.3531$ \\
$w_{g\gamma}^{D\,({\rm cen})}/w_{g\Gamma}^{D\,({\rm cen})}$ & $0.1123 \pm 0.0747\ (3.26\,(5)),\ 0.6596$ & $0.0827 \pm 0.0617\ (7.35\,(8)),\ 0.4998$ \\
$w_{g\gamma}^{E\,({\rm cen})}/w_{g\Gamma}$ & $4.7172 \pm 0.6937\ (4.42\,(5)),\ 0.4902$ & $3.0752 \pm 0.5012\ (16.40\,(8)),\ 0.0370$ \\
$w_{g\gamma}^{DB\,({\rm cen})}/w_{g\Gamma}$ & $2.4258 \pm 0.8563\ (3.59\,(5)),\ 0.6103$ & $1.9651 \pm 0.5698\ (15.23\,(8)),\ 0.0548$ \\
$w_{g\gamma}^{D\,({\rm cen})}/w_{g\Gamma}$ & $1.4921 \pm 0.7180\ (4.05\,(5)),\ 0.5418$ & $0.6159 \pm 0.4707\ (8.56\,(8)),\ 0.3811$ \\
\hline
\end{tabular}
\end{table*}

\bsp	%
\label{lastpage}
\end{document}